\documentclass[
    a4paper,
    nofootinbib,
    notitlepage,
    onecolumn,
    preprintnumbers,
    superscriptaddress
]{revtex4-1}

\usepackage{amsmath,amssymb}
\usepackage{braket}
\usepackage{graphicx}
\usepackage{hyperref}
\usepackage[%
    utf8
]{inputenc}
\usepackage{siunitx}
\usepackage[
    usenames,
    dvipsnames
]{xcolor}

\newcommand{\refeq}[1]{eq. (\ref{eq:#1})}

\newcommand{\refsec}[1]{section \ref{sec:#1}}

\newcommand{\dd}{\ensuremath{\textrm{d}}}
\newcommand{\GeV}{\si{\giga\electronvolt}}

\newcommand{\para}{\parallel}
\newcommand{\order}[1]{\ensuremath{\mathcal{O}\left(#1\right)}}

\begin{document}

\title{Prospects for disentangling long- and short-distance effects in the decays $B\to K^* \mu^+\mu^-$}
\author{Marcin Chrzaszcz}
\affiliation{European Organization for Nuclear Research (CERN), Geneva, Switzerland}
\affiliation{Universit\"at Z\"urich, Physik Institut, Winterthurer Strasse 190, 8057 Z\"urich, Switzerland}
\affiliation{Henryk Niewodniczanski Institute of Nuclear Physics Polish Academy of Sciences, Krak\'ow, Poland}
\author{Andrea Mauri}
\affiliation{Universit\"at Z\"urich, Physik Institut, Winterthurer Strasse 190, 8057 Z\"urich, Switzerland}
\author{Nicola Serra}
\affiliation{Universit\"at Z\"urich, Physik Institut, Winterthurer Strasse 190, 8057 Z\"urich, Switzerland}
\author{Rafael Silva Coutinho}
\affiliation{Universit\"at Z\"urich, Physik Institut, Winterthurer Strasse 190, 8057 Z\"urich, Switzerland}
\author{Danny van Dyk}
\affiliation{Universit\"at Z\"urich, Physik Institut, Winterthurer Strasse 190, 8057 Z\"urich, Switzerland}
\affiliation{Physik Department, Technische Universit\"at M\"unchen, James-Franck-Stra\ss{}e 1, 85748 Garching, Germany}

\preprint{TUM-HEP-1108/17}
\preprint{ZU-TH 25/17}

\begin{abstract}
	Theory uncertainties on non-local hadronic effects limit the New Physics 
	discovery potential of the rare decays $B\to K^*\mu^+\mu^-$.  We 
	investigate prospects to disentangle New Physics effects in the short-distance 
	coefficients from 	these effects. Our approach makes use of 
	an event-by-event amplitude analysis, and relies on the state of the art parametrisation 
	of the non-local contributions. 
	We find that non-standard effects in the short-distance coefficients can be 
	successfully disentangled from non-local hadronic effects. 
        The impact of the truncation on the parametrisation of non-local contributions to the Wilson 
        coefficients are for the first time systematically examined and prospects for its precise determination are discussed. 
        We find that physical observables are unaffected by these uncertainties.  
	Compared to other methods, our approach provides for a more precise extraction 
	of the angular observables from data.
\end{abstract}

\maketitle

\section{Introduction}
\label{sec:intro}

The sensitivity of the decay $B\to K^*\mu^+\mu^-$ to effects of beyond the
Standard Model (SM) physics is well known (see \emph{e.g.} \cite{Bediaga:2012py} for a
review). Consequently, this decay is the standard candle in indirect searches
for New Physics (NP) effects.  A recent analysis of this decay by the LHCb
collaboration has first established \cite{Aaij:2013qta} the so-called
$P'_5$ \cite{DescotesGenon:2012zf} ``anomaly'', \emph{i.e.} a deviation in measurements of the eponymous
observable from the SM predictions by $\sim 3\sigma$. Subsequent analyses by both LHCb
\cite{Aaij:2015oid} and Belle \cite{Wehle:2016yoi} further increased the
tension between the SM predictions and the data. Non-standard measurements
of the Lepton-Flavor-Universality (LFU) ratios in $b\to s\ell\ell$ processes
-- such as of $R_{K}$ and $R_{K^*}$~\cite{Hiller:2003js} by LHCb \cite{Aaij:2014ora,Aaij:2017vbb} -- 
suggest that a NP explanation of the $P'_5$ anomaly could simultaneously be LFU violating.\\

In their attempts to understand the anomalies, many phenomenological studies of
this decay strive to model-independently constrain the effects of New Physics.
This is usually achieved within the framework of an effective field theory.
Within the latter, a subset of the Wilson coefficients $C_i$ for the basis of
dimension-six operators $O_i$ are fitted from data. For the purpose of
this letter, we use the effective weak Lagrangian \cite{Bobeth:2012vn},
\begin{equation}
    \mathcal{L}^\text{eff} = \frac{4 G_F}{\sqrt{2}} V_{tb}^{} V_{ts}^* \left\lbrace
        \left[C_1 O_1^c + C_2 O_2^c\right]
        + \frac{\alpha_e}{4\pi}\left[C_7 O_7 + C_9 O_9 + C_{10} O_{10}\right]
    \right\rbrace + \order{\frac{V_{ub}^{} V_{us}^*}{V_{tb}^{} V_{ts}^*}, C_{3,\dots,6}, \alpha_s C_8}\,,
\end{equation}
with the current-current operators
\begin{align}
    O_1^c & = \left[\bar{s} \gamma^\mu P_L T^A c\right] \, \left[\bar{c} \gamma_\mu P_L T^A b\right]\,, &
    O_2^c & = \left[\bar{s} \gamma^\mu P_L c\right]     \, \left[\bar{c} \gamma_\mu P_L b\right]\,,
\end{align}
and radiative/semileptonic operators
\begin{align}
    O_7    & = \frac{m_b}{e} \left[\bar{s} \sigma^{\mu\nu} P_R b\right]\,F_{\mu\nu}\,, &
    O_9    & = \left[\bar{s} \gamma^\mu P_L b\right]\,\left[\bar\ell \gamma_\mu \ell\right]\,, &
    O_{10} & = \left[\bar{s} \gamma^\mu P_L b\right]\,\left[\bar\ell \gamma_\mu \gamma_5 \ell\right]\,.
\end{align}
Our conventions are the same as the ones of ref.\ \cite{Bobeth:2017vxj}, which
we follow closely.\\

A detriment to extracting the $C_i$, for $i=7,9,10$ from data is our lack of knowledge of the
hadronic matrix elements of the operators $O_i$. For local interactions, these
matrix elements are expressed in terms of hadronic form factors. The latter can
be accessed either from first principles through Lattice QCD simulations, or
from quark-hadron-duality arguments through QCD Light-Cone Sum Rules (LCSRs).
The matrix elements of non-local operators involving insertions of $O_{1,2}$, however, turn out to be more
difficult to determine, and have been the focus of much attention over the last
two decades. Presently, the largest systematic uncertainty in determinations of
the Wilson Coefficient $C_9$ arises from our lack of understanding of the
non-local hadronic matrix elements
\begin{equation}
    \mathcal{H}^\mu(q, k) \equiv \int d^4x \, e^{iqx} \bra{K^{*}(k)} \mathcal{T}\lbrace J_\text{e.m.}^\mu(x), \mathcal{O}_{bscc}(0)\rbrace \ket{\bar{B}(q + k)}\,
\end{equation}
where $J_\text{e.m.}$ denotes the electromagnetic current, and
\begin{equation}
    \mathcal{O}_{bscc} \equiv C_1 O_1^c + C_2 O_2^c\,
\end{equation}
represents the four quark operator involving two charm quark fields.  In the
above $k$ denotes the four-momentum of the final-state hadron, and $q$ describes the momentum
transfer to the virtual photon. It is convenient to decompose $\mathcal{H}^\mu$
into scalar-valued Lorentz-invariant quantities $\mathcal{H}_\lambda(q^2)$ as
in \cite{Bobeth:2017vxj}. Here $\lambda = 0,\perp,\para$ denotes the polarisation
state of the dilepton system.\\

The objects $\mathcal{H}_\lambda$ can be accessed in the limit of large
kaon energy in the $B$ rest frame, or equivalently for $q^2 \lesssim \textrm{a
few}\,\GeV^2 \ll m_b^2$ \cite{Beneke:2001at,Beneke:2004dp}. This QCD-improved
Factorisation (QCDF) approach has inspired a larger number of phenomenological
analyses.
However, all of these studies treat the off-shell contributions from the charm
pair as perturbative.
This treatment is known to receive substantial corrections
from soft-gluon emissions \cite{Khodjamirian:2010vf} off the charm loop, even for
the region $1\,\GeV^2 \leq q^2 \leq 6\,\GeV^2$ that is usually used for phenomenological
studies. It is therefore not surprising that the QCDF
calculations do not agree with the measured observables, for example in the purely
hadronic decays $B\to K^* \lbrace J/\psi, \psi(2S)\rbrace$.
An alternative to a theoretical determination of $\mathcal{H}$ are
data-driven analyses\footnote{%
    These analyses apply also to the decay $B\to K\mu^+\mu^-$, which has a reduced
    complexity compared to the decay $B\to K^*\mu^+\mu^-$ discussed in this
    letter.
}
\cite{Lyon:2014hpa,Ciuchini:2015qxb,Aaij:2016cbx,Ciuchini:2017mik}. Some of
these analyses show promise in fitting a resonance model to the $q^2$ spectrum.
Others determine the non-local contributions below the $J/\psi$ from data.
Both approached therefore give access to \emph{model-dependent} determinations of the WC
$C_9$. In addition, information on the model parameters are only available
a-posteriori, which precludes genuine SM predictions of the observables.\\

Both drawbacks, the perturbative treatment of the charm quarks at timelike $q^2$,
and the model assumptions in the parametrisation of the $q^2$ spectrum have recently been overcome
through a parametrisation that is valid for $-7 \lesssim q^2 \leq M_{\psi(2S)}^2$ \cite{Bobeth:2017vxj}.
In that study, pseudo observables based on the theoretical predictions are generated at $q^2 < 0$,
for which one expects rapid convergence of the Light-Cone OPE \cite{Khodjamirian:2010vf}. In addition,
the residues of the scalar valued correlators $\mathcal{H}_\lambda(q^2)$ can be constrained
from experimental results on the branching ratios and angular distribution of the hadronic
decays $B\to \psi_n K^*$, where $\psi_n = J/\psi, \psi(2S)$. Last but not least, expressing
the ratio of the $\mathcal{H}_\lambda$ over their corresponding (local) form factors
$\mathcal{F}_\lambda$ in combination with an expansion in the conformal variable $z$ ensures
the correct analytic behaviour. Since we follow the results of reference \cite{Bobeth:2017vxj}
closely, the parametrisation used for the correlators $\mathcal{H}_\lambda(q^2)$ does not
reproduce the physical light-hadron cut starting at $q^2 = 4 M_\pi^2$. However, it has been
argued that this cut is suppressed \cite{Bobeth:2017vxj}, and we will discuss -- in parts -- its numerical
impact and the possible model bias that the lack of the cut introduces in \refsec{methodology-results}.\\

Given this recent progress on the non-local matrix elements we now aim to study
the possibility of applying the $z$ expansion to future experimental analyses:
First, we want to establish to which extent information concerning the non-local matrix
elements can be inferred from experimental data of the semileptonic decay. 
In order to
maximise the sensitivity to the parameters, we will focus on an extended unbinned
analysis of the data; see \cite{Hurth:2017sqw} for a related study of unbinned
analyses with focus on the $K\pi$ S-wave background.
Second, we investigate the convergency of the series expansion at different order of $z$
and what is the residual model-bias introduced by truncations.
Finally, we want to establish the smallest amount of theoretical inputs necessary
to find evidence for new phenomena in quark-flavor physics in a single $b\to s\ell\ell$
process and through a single measurement.

\section{Preliminaries}
\label{sec:preliminaries}

Assuming on-shell $K^*$ dominance, the decay $B\to K^*(\to K \pi)\mu^+\mu^-$ involves
four kinematic variables: the dimuon mass square $q^2$, as well as two helicity angles
within the $\mu^+\mu^-$ and $K \pi$ decay planes, respectively, and the azimuthal angle between
the planes (see \cite{Kruger:1999xa,Kruger:2005ep} and subsequent publications). The Probability
Density Function (PDF) for this decay gives rise to a larger number of angular observables,
which can be used to extract information
on the short-distance physics. Here, we do not use these angular observables directly, but rather
use the angular information of the signal PDF for $B\to K^*(\to K\pi)\mu^+\mu^-$ decays in its
entirety.

We work with two signal PDFs: PDF$_1$ and PDF$_2$, defined as
\begin{equation}
    \mathrm{PDF}_i \equiv \frac{1}{\Gamma_i}\,\frac{\dd^4\Gamma}{\dd q^2\, \dd^3\Omega}\,,
    \qquad
    \text{with}\quad
    \Gamma_i \equiv \int_{q^2 \in Q_i} \dd q^2\, \frac{\dd \Gamma}{\dd q^2}\,.
\end{equation}
Here the four-differential rate is a sesquilinear form of the set of transversity amplitudes
$A_{\lambda}(q^2)$, with polarisation states $\lambda=\perp,\para,0,t$ \cite{Bobeth:2012vn}. For later discussion,
it is instrumental to understand that the amplitudes with $\lambda = \perp,\para,0$ can (in the SM) be
written as \cite{Bobeth:2017vxj}
\begin{equation}
    \mathcal{A}_\lambda^{L,R}
    = \mathcal{N}_\lambda\ \bigg\{ 
        (C_9 \mp C_{10}) \mathcal{F}_\lambda(q^2) \\
        + \frac{2m_b M_B}{q^2} \bigg[ C_7 \mathcal{F}_\lambda^{T}(q^2)
        - 16\pi^2 \frac{M_B}{m_b} \mathcal{H}_\lambda(q^2) \bigg]
        \bigg\} + \order{C_{3,\dots,6,}, \alpha_s C_8, \big|V_{ub} V_{us}\big|}\,.
\end{equation}
In the above, the functions $\mathcal{F}_\lambda^{(T)}(q^2)$ stand for (linear combinations)
of local form factors (FF), while the non-local matrix elements $\mathcal{H}_\lambda(q^2)$ have
been introduced in \refsec{intro}.

The PDFs describe the combined $q^2$ and full-angular distribution of the decay in two kinematic
regions:
\begin{equation}
\begin{aligned}
    Q_1 &: \quad 1.1\,\GeV^2  \leq q^2 \leq 9.0\,\GeV^2\,, \\
    Q_2 &: \quad 10.0\,\GeV^2  \leq q^2 \leq 13.0\,\GeV^2\,.
\end{aligned}
\label{eq:def-bins}
\end{equation}
Note that we impose no constraints on the angular support in either of the PDFs.
For a definition of $\dd^4\Gamma / (\dd q^2\, \dd^3\Omega)$ we refer to \cite{Bobeth:2008ij,Altmannshofer:2008dz}
and references therein.\\

We produce toy ensembles using the central values of the input parameters
$\lbrace \alpha_j \rbrace \equiv \lbrace \alpha_k^{(\lambda)}\rbrace \cup %
\lbrace \alpha_l^{(F)} \rbrace \cup \lbrace \alpha_m^{(\text{CKM})} \rbrace$,
where the individual parameter sets are:
\begin{itemize}
    \item the correlator parameters $\lbrace \alpha_k^{(\lambda)}\rbrace$ for each polarisation
        $\lambda=\perp,\para,0$ as specified in \cite{Bobeth:2017vxj}, corresponding to a nominal truncation at $z^2$;
    \item the form factor parameters $\lbrace \alpha_l^{(F)}\rbrace$ for form factors
        $F=V,A_{0,\dots,2},T_{1,\dots,3}$ as determined from a combined fit to LCSR and lattice QCD results
        in \cite{Straub:2015ica}, but with twice the stated uncertainty;
    \item and the CKM Wolfenstein parameters $\lbrace \alpha_m^{(\text{CKM})}\rbrace%
        \equiv \lbrace \lambda, A, \bar{\rho}, \bar{\eta}\rbrace$,
        as obtained from a tree-level analysis of the unitarity triangle \cite{Bona:2006ah}.
\end{itemize}

Toy ensembles are either labelled as ``SM'', in which case we fix all
Wilson Coefficients (WCs) to their SM values; or ``Benchmark Point'' (``BMP''), in which case
the WC $C_9$ is shifted by $-1$ from its SM value. Each toy ensemble consists
of $N \equiv N_1 + N_2$ toy events, which are drawn from the combined log-PDF
$\ln \mathrm{PDF} \equiv \ln \mathrm{PDF}_1 + \ln \mathrm{PDF}_2$ of the aforementioned two  $q^2$ regions (see \refeq{def-bins}).
The total number of events $N$ is varied to
explore the sensitivity for present and future experiments.

In all cases under study, we maximise the total log-likelihood $\ln \mathcal{L}_\text{tot} \equiv \ln
\mathcal{L}_{1} + \ln\mathcal{L}_{2} + \ln \mathcal{L}_{\mathcal{B}}$ with
respect to the nuisance parameters $\lbrace \alpha_j \rbrace$ and additionally
-- for New Physics fits -- the WCs $C_9$ and (in some cases) $C_{10}$. Here $\mathcal{L}_\text{1,2}$ are
unbinned likelihoods of $N_i$ toy events $x_{n,i} \sim \mathrm{PDF}_i$,
\begin{equation}
    \ln \mathcal{L}_i\left(\lbrace \alpha_j\rbrace\right) \equiv \sum_{n}^{N_i} \mathrm{PDF}_i(x_{n,i}\,|\, \lbrace \alpha_j\rbrace)\,.
\end{equation}
The last term, $\mathcal{L}_\mathcal{B}$, incorporates two Poissonian terms for the
integrated branching ratios of the decay in the kinematics regions $Q_1$ and $Q_2$, respectively. \\

We then perform a series of frequentist fits to determine the viability of the approach, and
to determine uncertainty intervals for the various parameters.
In our nominal fits up to $z^2$, we float the full set of $39$ nuisance parameters
$\lbrace \alpha_j \rbrace$, with Gaussian constraints as described above. Exception
to this are marked appropriately in the text.
All quoted $68\%$ confidence level intervals are determined from profile likelihoods. 
For our studies we are using the same setup as
in \cite{Bobeth:2017vxj}, however independently implemented and cross-checked
against the EOS software \cite{EOS-web}.\\

\begin{table}[t]
\begin{tabular}{c|ccccc}
    &   {LHCb Run I} & {LHCb Run II} & {LHCb Upgrade [$50$ fb$^{-1}$]}  & {Belle II [$50$ ab$^{-1}$]}\\
\hline
$N$ &   1850         & 6,900     & 62,000             & 6,100
\end{tabular}
\caption{
    Number of produced pseudo events per toy experiment.
}
\label{tab:number-of-events}
\end{table}
The LHCb experiment has already observed $969$ and $330$ $B^{0} \to K^{*0}\mu^+\mu^-$ events in the bins
$1.1\,\GeV^2 \leq q^2 \leq 8.0\,\GeV^2$ and $11.0\,\GeV^2 \leq q^2 \leq
12.5\,\GeV^2$, respectively~\cite{Aaij:2015oid}.  Extrapolating in $q^2$ to the
larger bin widths as defined in \refeq{def-bins}, we fix $N_\text{LHCb-Run I}
\equiv 1850$. 
In the same fashion we obtain $N_{\textrm{Belle}} = 56$.
In order to study the sensitivities for future data sets, we
extrapolate the number of events by scaling the luminosities and $b\bar{b}$
production cross section $\sigma_{b\bar{b}}(s)$, where $s$ denotes the designed
centre-of-mass energy of the \textit{b}-quark pair.  For the LHCb experiment we
use $\sigma_{b\bar{b}} \propto \sqrt{s}$, while for Belle-II we use
$\sigma_{b\bar{b}} \propto s$. The exact numbers of simulated events for each
experiment are listed in table \ref{tab:number-of-events}.\\

Modelling of both the detector resolution or detection efficiency is hardly possible
without access to (non-public) information of the current $B$ physics experiments Belle (II)
and LHCb. We therefore assume perfect resolution and efficiency in our studies herefrom out - unless otherwise stated.
As a consequence, all of our following results should be understood as upper
bounds on the possible sensitivity of any future experimental analysis
following our suggestions.\\

Note that we do not study the hadronic uncertainties in the context of free floating
parameters for further WCs%
\footnote{%
    We thank S\'ebastien Descotes-Genon for raising this question in private communications.
}
beside $C_9$ and $C_{10}$, since the remaining WCs of semileptonic operators can be disentangled from $C_9$
as various global analyses of $b\to s\ell\ell$ processes have shown; see \emph{e.g.}
\cite{%
Descotes-Genon:2013wba,%
Altmannshofer:2013foa,%
Beaujean:2013soa,%
Hurth:2013ssa,%
Beaujean:2015gba%
}.
Moreover, the hadronic non-local effects can always be attributed to $q^2$-dependent shifts
of the WCs $C_9$ and $C_{9'}$ (the chirality-flipped counterpart to $C_9$).
Sizeable shifts to $C_{9'}$ require NP contributions of non-SM chirality in the operators $O_{1'(2')}^{(c)}$,
and are not further discussed here; see also \cite{Jager:2017gal} for a related
discussion in the presence of (pseudo)scalar four-quark operators.
Consequently, we are convinced that it suffices to demonstrate the separability of
hadronic and NP contributions to $C_9$ as we set out to do in this article.

\section{Initial study}
\label{sec:methodology-results}

Our main questions concerning the prospects of future analyses pertain to disentangling
the hadronic effects of the hadronic correlators from NP effects in the Wilson coefficient
$C_9$. 

These questions are:
\begin{enumerate}
    \item[A] Can the parameters $\lbrace \alpha_k^{(\lambda)}\rbrace$ describing the
        non-local charm contributions be extracted from semileptonic data only (\emph{i.e.}, without
        theory input through Gaussian constraints)?
    \item[B] The present theory results for the parameters $\lbrace \alpha_k^{(\lambda)} \rbrace$ of
        the hadronic correlator assume stable charmonia $J/\psi$ and $\psi(2S)$, i.e., vanishing width
        for these states. Does neglecting their finite widths introduce a
        numerically relevant systematic bias in the extraction of the nuisance parameters?
    \item[C] What residual model-bias is introduced by cutting off the series expansion of the
        $\mathcal{H}_\lambda(q^2)$ at the power of $z^2$?
\end{enumerate}

\subsection{Determining the $\alpha_\lambda^{(k\leq 2)}$ parameters without theory constraints}
\label{sec:methodology-results:no-constraints}

Prior theoretical knowledge can be used in the fits through Gaussian constraints on the nuisance
parameters $\alpha_\lambda^{(k)}$ describing
the non-local correlator. However, these constraints can only be produced for parametrisations
with a truncation at $z^2$ \cite{Bobeth:2017vxj}.
Here we investigate if, in principle, one could abstain from using Gaussian constraints on these parameters,
and instead determine them only from data of the semileptonic decay.\\

In order to answer this question, we perform our analyses in two ways.  For our
first analysis, we use the constraints provided in \cite{Bobeth:2017vxj}, which
are based on theory calculations in the negative $q^2$ region, as well as
experimental measurements of the angular distributions of the decays $B\to K^*
J/\psi$ and $B\to K^*\psi(2S)$. These constraints are included in form of a multivariate
Gaussian distribution with correlation information taken into account.  Our second analysis
does not use either source of information as a constraint on the parameters
$\lbrace \alpha_\lambda^{(k)}\rbrace$, and floats them instead within the range
$\alpha_\lambda^{(k)} \in [-10^{-2}, +10^{-2}]$.\\

A series of 500 toy data sets have been produced with the BMP
scenario, and then fitted with and without usage of the constraints. From our
toy analyes we conclude the following:
\begin{enumerate}
    \item The analysis with Gaussian constraints on the parameters $\lbrace \alpha_\lambda^{(k)}\rbrace$
        is able to extract additional information on the hadronic correlators from the fit,
        \textit{i.e.}, the obtained uncertainties are smaller than the corresponding Gaussian constraints.
        The uncertainties on the hadronic parameters scale by a factor between 0.5 and 0.8 for the expected statistics of the LHCb Run II.
    \item When removing the constraints, we find that the fit still converges, and that
          we are able to disentangle hadronic effects from NP in $C_{9}$.
          Our estimator for $C_9$ is unbiased for a large number of events $\sim 30$k.
          For the expected statistics of the LHCb Run II 
          the uncertainties obtained from the fit on the parameters $\lbrace \alpha_\lambda^{(k)}\rbrace$
          are found to be comparable to the ones from the prior predictions \cite{Bobeth:2017vxj}.
\end{enumerate}

\subsection{Finite-width effects}

The constraints for the parameters $\lbrace \alpha_k \rbrace$ are based
\cite{Bobeth:2017vxj} on theoretical results (at negative $q^2$) and the
residues of the $\psi = \lbrace J/\psi, \psi(2S)\rbrace$ poles of the
correlation functions $\mathcal{H}_\lambda(q^2)$ in the complex $q^2$ plane.
Information on the residues can be obtained from experimental results
for the angular distribution of the decays $B\to K^*(\to K\pi) \psi(\to \mu^+\mu^-)$
in small mass windows around the respective $\psi$ masses. In reference
\cite{Bobeth:2017vxj}, the narrow charmonia have been assumed to be stable to
simplify the discussion. As a consequence, the poles of the proposed
parametrisation are located on the real $q^2$ axis. However, the physical poles
are shifted below the real axis by finite-width effects. Moreover, the shift is directly connected to
the $q^2$ shape of the resonances.\\

In order to study possible bias introduced by neglecting the width of the
narrow charmonia, we study an ensemble of $1$k toy analyses
corresponding to $N_\text{LHCb Upgrade} = 62$k events each. We produce the toy
events for each fit by using the SM scenario.
For each toy analysis we perform two fits to the toy
data: one with the nominal PDF, and one for which we modify the PDF such that
the poles corresponding to the two narrow charmonium states are shifted below
the real $q^2$-axis by $i M_\psi \Gamma_\psi$. Note that since this shift is not
$q^2$ dependent, the induced imaginary part does not vanish for $q^2
< 4 M_\pi^2$. Consequently, our fit PDF does not respect unitarity. However,
since it is only used to model possible fit bias at $q^2 \geq 1\,\GeV^2 \gg 4 M_\pi^2$,
this does not pose a problem here.\footnote{%
    Effects of non-vanishing width very close to or on the $J/\psi$, $\psi(2S)$ resonances, as well as for $q^2 \leq 0
    $ are tightly related to the theory prior and therefore not studied here.
}\\

We carry out both fits, with the nominal and the modified PDF \emph{without the
usage of any Gaussian constraints on the parameters $\lbrace \alpha_k
\rbrace$}. We find that the results of the fits to toys with and without the
width effects are indistinguishable even for an ensemble corresponding to the
LHCb Upgrade. As such, the bias introduced by neglecting the finite width \emph{in the
semileptonic regions} $Q_1$ and $Q_2$ will not play a relevant role for any of
the upcoming data sets.

\subsection{Model bias of and sensitivity to higher orders in $z$}
\label{sec:methodology-results:higher-orders}

In the analysis \cite{Bobeth:2017vxj} the truncation order was chosen as $K=2$, in order to ensure that
a-priori predictions (i.e., genuine SM predictions without using $B\to K^*\mu^+\mu^-$ data) can be made. In the applications discussed here
we are not bound to using the priors presented in \cite{Bobeth:2017vxj} as Gaussian constraints;
see \refsec{methodology-results:no-constraints}. Thus, we can explore the sensitivity to coefficients
that enter with $z^3$ or even higher powers in $z$. Including these in the experimental analysis
has the potential to reduce the model-dependence of the results on $C_9$ due to the $z$
truncation.\\

We choose to probe the sensitivity to coefficients of order $z^3$ as follows: We first produce $4$k
toy analyses with data sets corresponding to the LHCb Run II luminosity.
For each study, the pseudo events follow from the BMP scenario.
In addition, we introduce the coefficients
$\lbrace \alpha_3^{(\lambda)}\rbrace$ for the higher order terms in the correlators
$\mathcal{H}_\lambda(q^2)$. We produce toys for $\alpha_3^{(\lambda)} = 0$ $\forall$ $\lambda=\perp,\para,0$.
However, in the fit we let the $\lbrace \alpha_3^{(\lambda)}\rbrace$ float
freely within $[-0.1, +0.1]$, and determine their $68\%$ CL intervals.
We find
\begin{equation}
\begin{aligned}
    \sigma\left(\operatorname*{Re}{\alpha_3^{(\lambda)}}\right)
        & = 5\cdot 10^{-3}\,,  &
    \sigma\left(\operatorname*{Im}{\alpha_3^{(\lambda)}}\right)
        & = 6\cdot 10^{-3}\,.
\end{aligned}
\end{equation}
Consequently, we find no sensitivity to coefficients which are smaller in magnitude than $5\cdot
10^{-3}$.

We also investigate if the value of $C_9$ obtained from a fit to order $z^3$ is fully compatible
with the fit to order $z^2$. Our analysis yields
\begin{equation}
	C_9\big|_\text{$z^3$ fit} - C_9\big|_\text{$z^2$ fit} = {0.17}
\end{equation}
and
\begin{equation}
	\sigma(C_9)\big|_\text{$z^2$ fit} = {0.19} \qquad \text{and} \qquad \sigma(C_9)\big|_\text{$z^3$ fit} = {0.69}\,.
\end{equation}
Our findings can be summarised as follows:
\begin{itemize}
    \item The impact of $z^3$ terms on the extraction of $C_9$ amounts to a shift by (formally)
        less than one standard deviation of the fit to order $z^2$, when considering data set sizes
        up to the LHCb Run-II size.
    \item The model-dependence of the fit to order $z^3$ is large in the absence of any theory constraints
        on the parameters $\lbrace \alpha_\lambda^{(k)}\rbrace$.
\end{itemize}
We conclude that a small model-dependence can only be achieved by using more information than only
the experimental data of the semileptonic decay $B\to K^*\mu^+\mu^-$. However, when using order $z^3$
or higher, the current set of theoretical and experimental constraints does not allow for a staged approach
\cite{Bobeth:2017vxj}, in which the posterior of a $C_9$-agnostic fit is used as a prior for the NP fit.
We see three possibilities to overcome this problem:
\begin{enumerate}
    \item through a \emph{staged} analysis that uses theoretical predictions beyond what has been discussed in
    \cite{Bobeth:2017vxj} (e.g. lattice QCD calculations of the hadronic correlator on or in between the narrow
    charmonium resonances);
    \item through a \emph{simultaneous} analysis of the semimuonic \emph{and} semielectronic decays, in which
    the parameters for the non-local matrix elements are shared (as shown in~\cite{Mauri:LFU});
    \item through a \emph{combined} analysis of the theory predictions at negative $q^2$, the measurements
    of the hadronic decays $B\to K^*\lbrace J/\psi, \psi(2S)\rbrace$, \emph{and} the measurements of
    the semileptonic decays. This approach is investigated in the next section.
\end{enumerate}

\section{Combined unbinned analysis}
In this section we extend our previous analysis by performing a combined unbinned
fit of the theory predictions at negative $q^2$, the measurements
of the hadronic decays $B\to K^*\lbrace J/\psi, \psi(2S)\rbrace$, and the measurements of
the semileptonic decays.
We investigate the following points:
\begin{enumerate}
    \item[A] Can the residual model-bias seen in section~\ref{sec:methodology-results:higher-orders} be reduced by including theoretical predictions and
        measurements of the hadronic decays $B\to K^*\lbrace J/\psi, \psi(2S)\rbrace$?
    \item[B] What is the individual impact of the theoretical predictions and the hadronic decays inputs? 
    \item[C] In this combined analysis, what model bias is introduced by the truncation assumption at the production level? 
    \item[D] What are the prospects for a simultaneous fit to the WCs $\mathcal{C}^{\textrm{NP}}_9$ and $\mathcal{C}^{\textrm{NP}}_{10}$?
    \item[E] What is the gain in precision when determining the shape of the $q^2$-differential angular observables compared to binned analyses?
\end{enumerate}

\subsection{Combined fit to theory points at $q^2 < 0$, hadronic and semileptonic decays}
\label{sec:CombinedAnalysis}

In order to perform a combined fit to all the available (theoretical and experimental) information,
we extend the current framework to include the predictions on the hadronic correlator calculated for points 
at negative $q^2$ and the two sets of pseudo-observables (three magnitudes and two relative phases
for each of the $J/\psi$ and $\psi(2S)$ resonances), as in~\cite{Bobeth:2017vxj}.
Both contributions are included in the fit as multivariate Gaussians of the relevant pseudo-observables.
For the production of the ensembles the corresponding central values are shifted to match their 
predictions given a certain set of parameters $\lbrace \alpha_\lambda^{(k)}\rbrace$, while the 
uncertainty is scaled to keep the relative error constant. 
Unless stated otherwise, ensembles are drawn from the BMP scenario,  
\textit{i.e.} all the coefficients $\lbrace \alpha_\lambda^{(k)}\rbrace$ of order higher than $z^2$ are set to zero.

We explore the model-bias introduced by the truncation of the series by repeating the fit with different 
truncation orders for the $z$ expansion.
Two sets of toy analyses with 500 ensembles each are produced, with data sets corresponding to the 
LHCb Run II and LHCb Upgrade [50 fb$^{-1}$] luminosity, respectively.

In Table~\ref{tab:pullC9} we report the resulting sensitivity to the $\mathcal{C}^{\textrm{NP}}_9$ obtained for
$z^2$, $z^3$, $z^4$ and $z^5$ fits.
Our findings can be summarised as follows:
\begin{itemize}
    \item The uncertainty on $\mathcal{C}^{\textrm{NP}}_9$ roughly doubles moving from
    $z^2$ fits to $z^3$ fits, for both statistics under consideration.  
    \item For the dataset corresponding to the expected statistics at the LHCb Upgrade [50 fb$^{-1}$], the 
    uncertainty on $\mathcal{C}^{\textrm{NP}}_9$ slightly improves for orders higher than $z^3$.
    \item For the dataset corresponding to the expected statistics at the LHCb Run II we observe the saturation of the uncertainty 
    at the orders $z^3$ and $z^4$. However, for the fit with $z^5$ the uncertainty starts to increase, pointing to a statistical limitation.
\end{itemize}

\begin{figure}[tbh!]
\includegraphics[width=.49\textwidth]{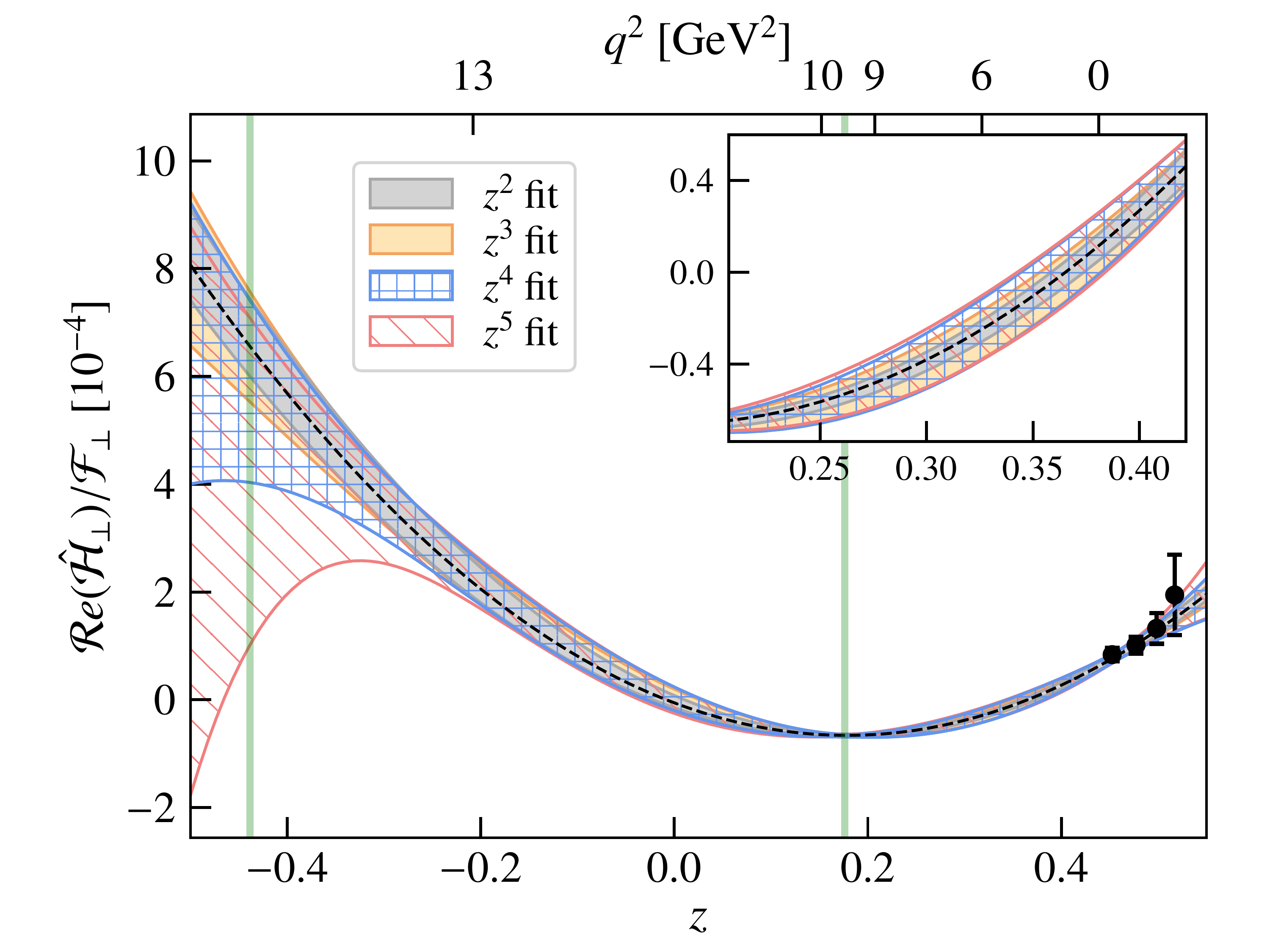} 
\includegraphics[width=.49\textwidth]{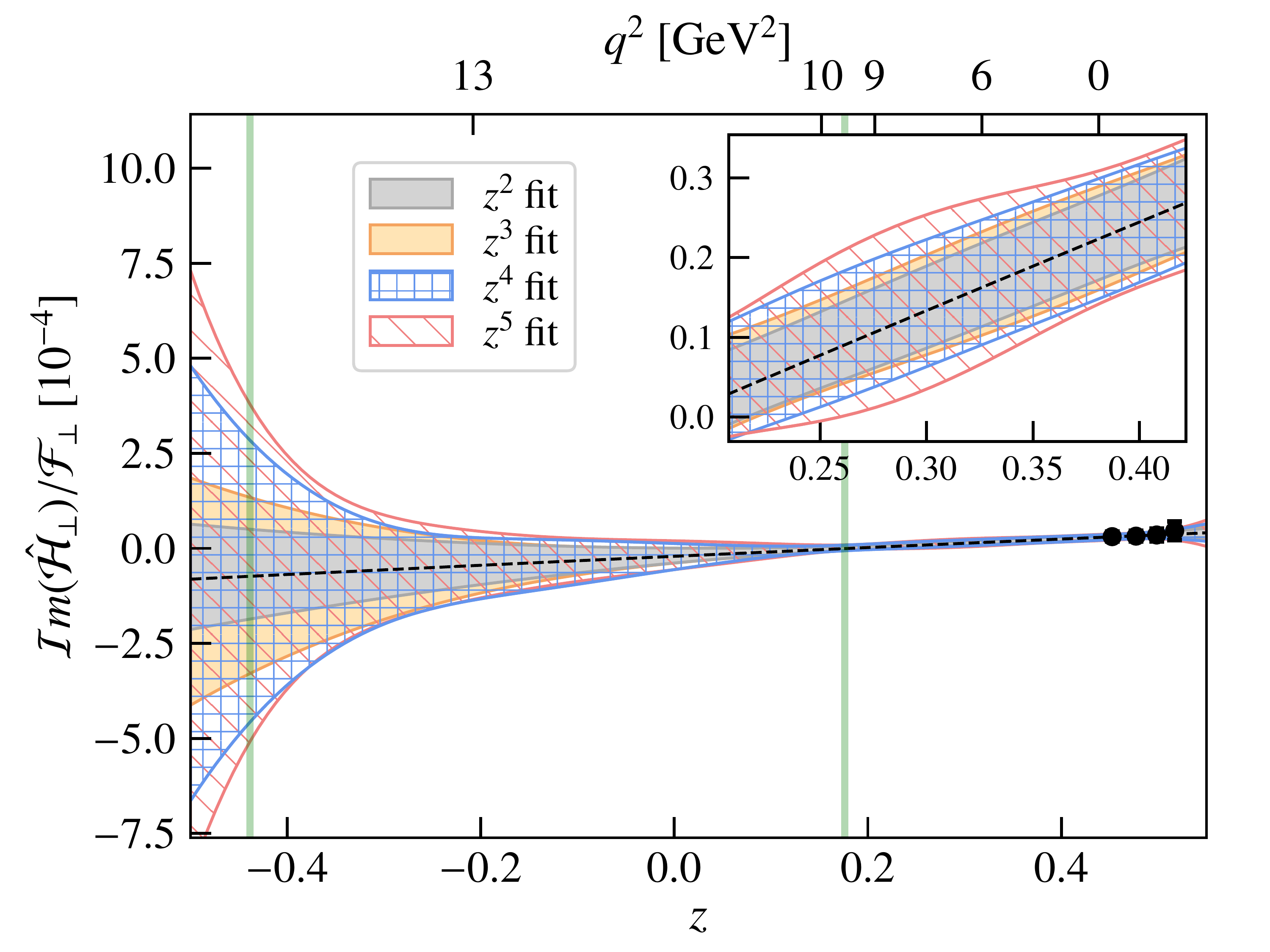} \\
\vspace{5mm}
\includegraphics[width=.49\textwidth]{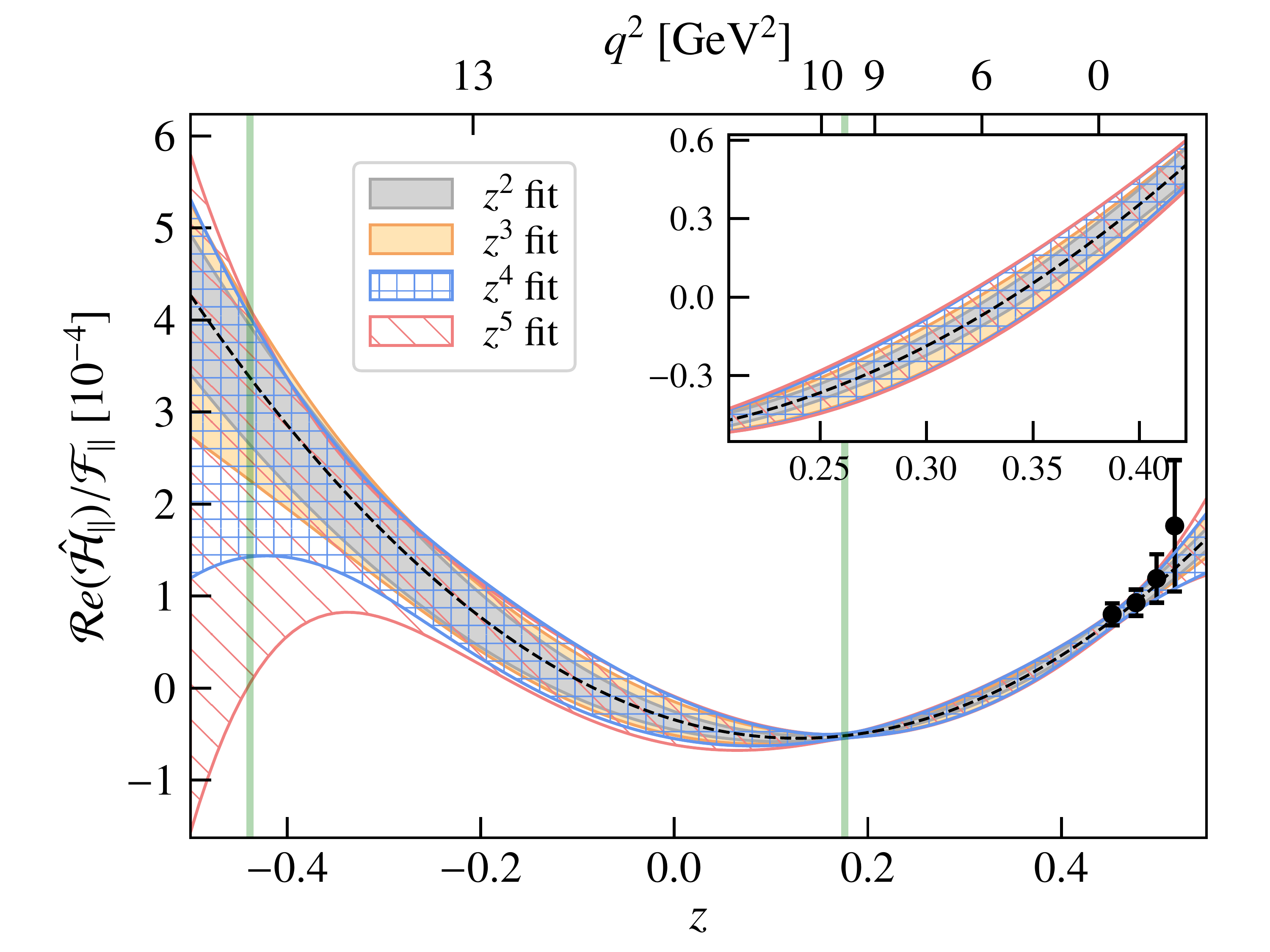} 
\includegraphics[width=.49\textwidth]{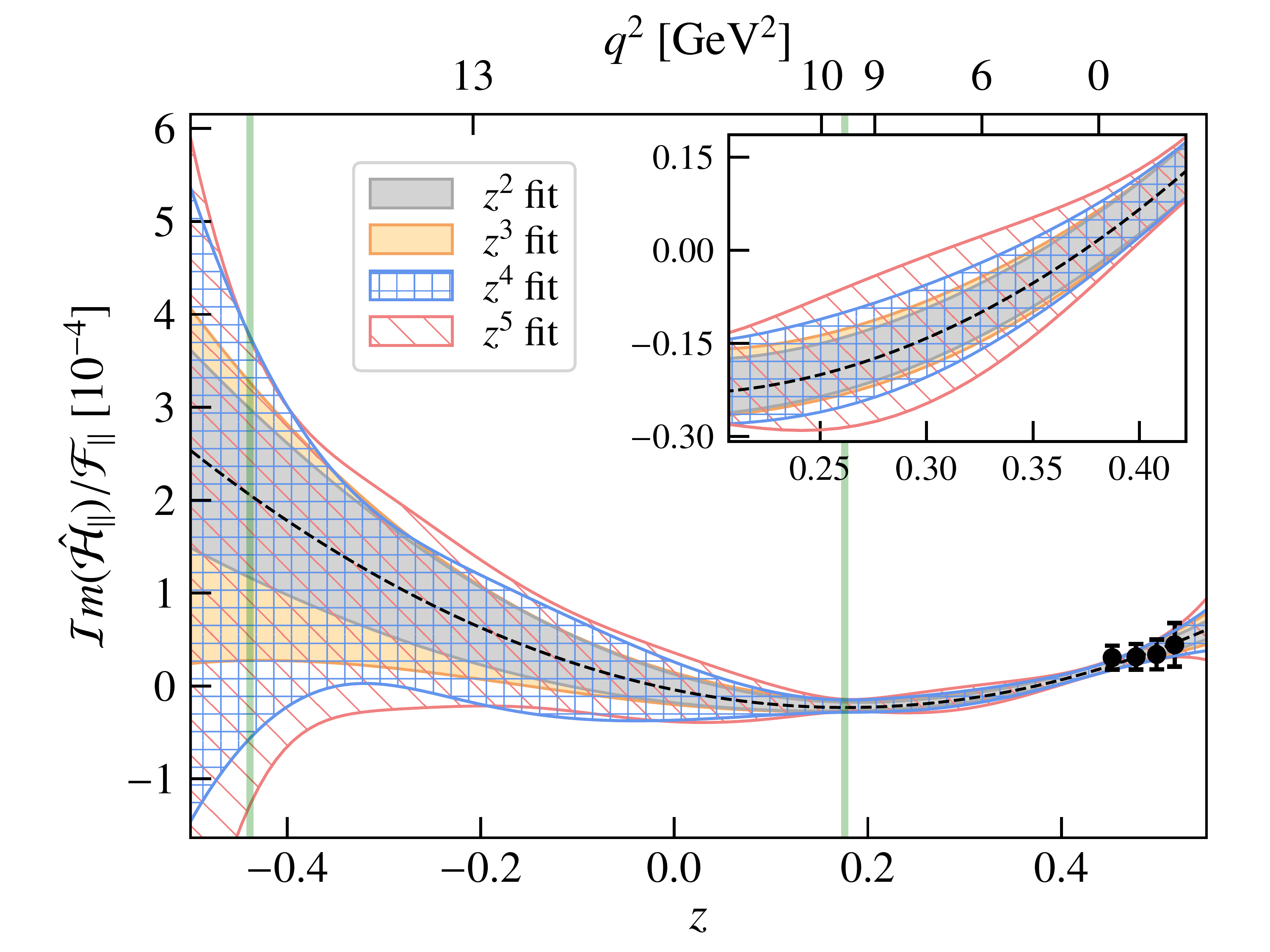} \\
\vspace{5mm}
\includegraphics[width=.49\textwidth]{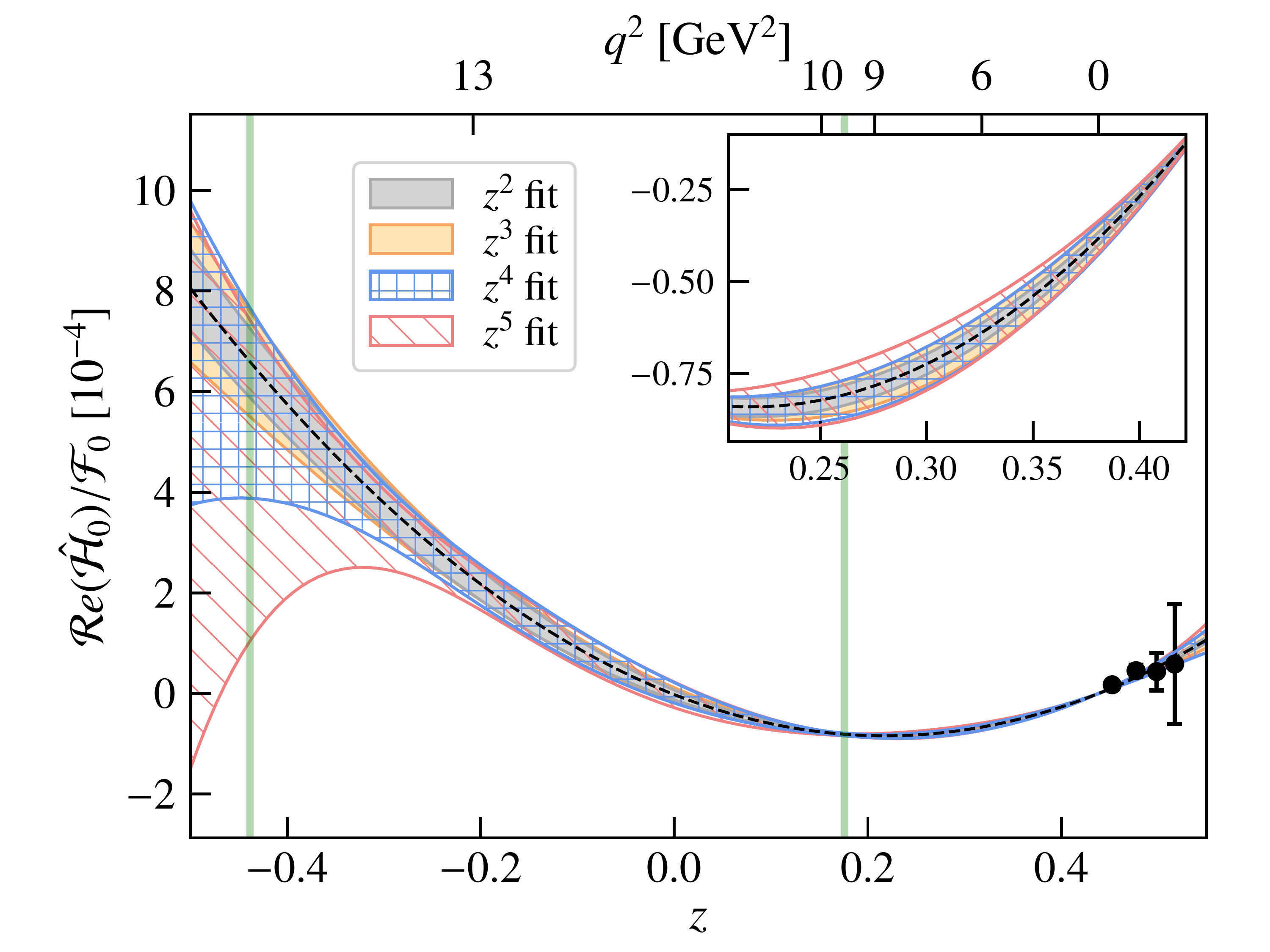} 
\includegraphics[width=.49\textwidth]{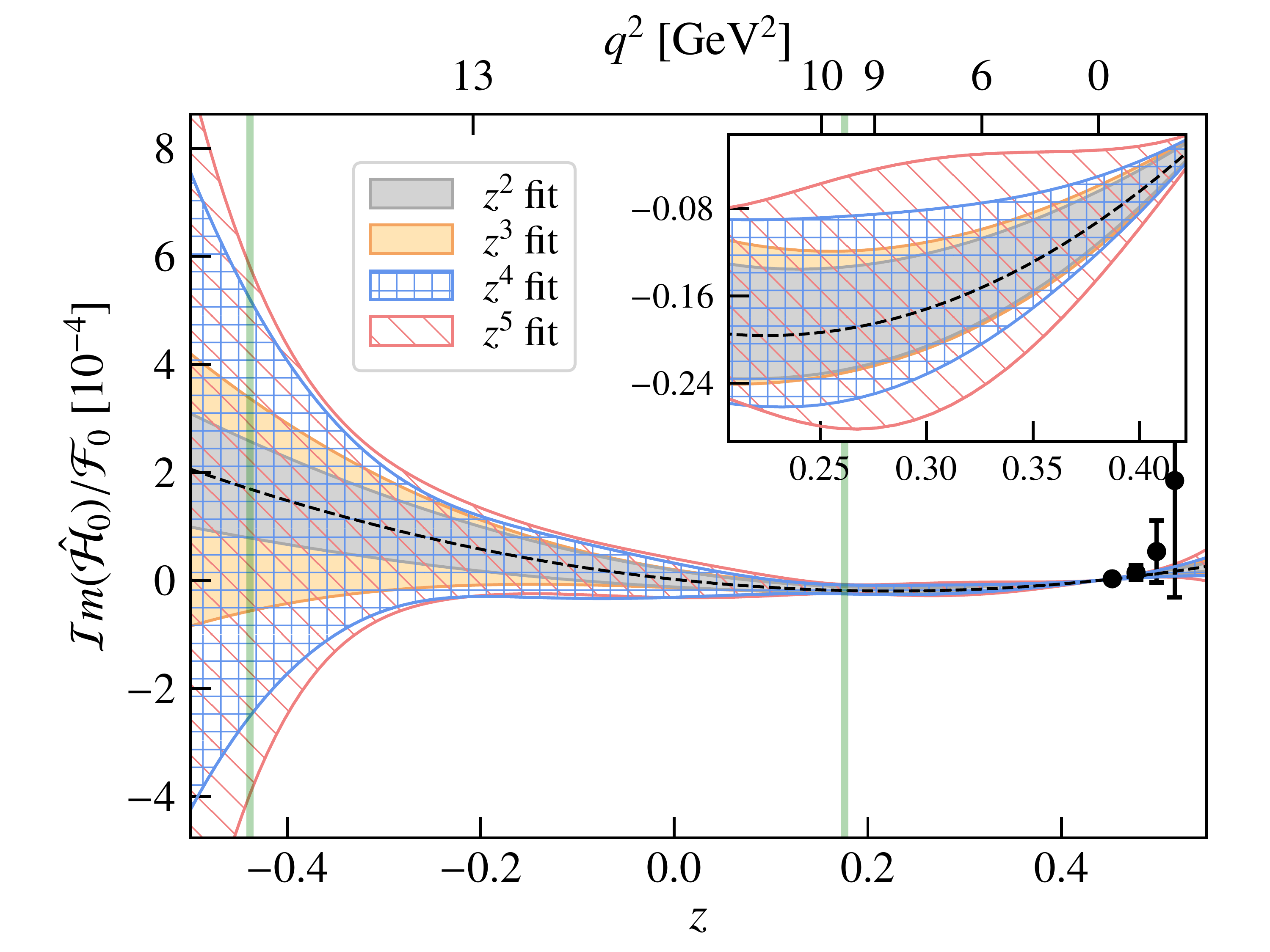} \\
\caption{%
    Results of the fits for the ratio Re $\hat{\mathcal{H}}_\lambda(z)/\mathcal{F}_\lambda(z)$ obtained with different orders 
    of the expansion for the BMP scenario and the expected statistics at LHCb Upgrade [50 fb$^{-1}$].
    The vertical bands correspond to the $J/\psi$ and $\psi(2S)$ regions; 
    the points to the theoretical inputs at negative $q^2$.
    The top right box of each plot zooms in the $q^2$ range between 1.1 and 9.0 GeV$^2$. 
    \label{fig:H}
}
\end{figure}

The observed distributions of the hadronic correlator $\mathcal{H}_\lambda(q^2)/\mathcal{F}_\lambda(q^2)$ for different 
orders in $z$ are shown in Fig.\ref{fig:H}, corresponding to the statistics expected at the LHCb Upgrade [50 fb$^{-1}$].
In all cases, the uncertainty drastically increases for higher orders in the $\psi(2S)$ window.
For the regions $Q_1$ and $Q_2$, however, we find that the behaviour observed for $\mathcal{C}^{\textrm{NP}}_9$
is reflected in the real part of the hadronic correlator, \textit{i.e.} the uncertainty saturates for orders higher than $z^3$.

\begin{table}
  \begin{center}
    \begin{tabular}{ c c c c c }
      \hline
      													  &      						 \multicolumn{2}{c}{LHCb Run2}									&  				   \multicolumn{2}{c}{LHCb Upgrade [50 fb$^{-1}$]}   				\\
      													  &      Re$\,\mathcal{C}_9^{\textnormal{NP}}$ mean      &      Re$\,\mathcal{C}_9^{\textnormal{NP}}$ sigma  &      Re$\,\mathcal{C}_9^{\textnormal{NP}}$ mean      &      Re$\,\mathcal{C}_9^{\textnormal{NP}}$ sigma    	 \\
      \hline
      $z^2$ fit    &     		-0.966 $\pm$ 0.006  		   &   		   0.120 $\pm$ 0.004    		&   		  -0.996 $\pm$ 0.003    		&    			0.060 $\pm$ 0.002      		\\ 
      $z^3$ fit    &    		 -0.991 $\pm$ 0.011  	  &      			 0.217 $\pm$ 0.008   	 &  		   -1.015 $\pm$ 0.006   		 &    			 0.124 $\pm$ 0.004     		 \\ 
      $z^4$ fit    &    		 -1.022 $\pm$ 0.011  	   &     			 0.229 $\pm$ 0.008     &    			-1.012 $\pm$ 0.007   	 	&     			 0.146 $\pm$ 0.005     	 	\\ 
      $z^5$ fit    &   		  -0.942 $\pm$ 0.016 	   &       		0.293 $\pm$ 0.011    	&    			 -0.983 $\pm$ 0.008    		 &     		0.157 $\pm$ 0.006     		 \\ 
      \hline
    \end{tabular}
    \caption{Expected central value and uncertainties for the $\mathcal{C}^{\textrm{NP}}_9$ observable obtained from $z^2$, $z^3$, $z^4$ and $z^5$ fits for the BMP scenario with Re$\,\mathcal{C}_9^{\textnormal{NP}}=-1$.}
    \label{tab:pullC9}
  \end{center}
\end{table}

\subsection{Exploring the impact of the inputs from theory and hadronic decays}

We further investigate the impact of the additional pseudo-observables introduced in the combined fit,
to distinguish the benefits obtained from either of the two inputs.
In particular, we investigate the following questions:
\begin{itemize}
    \item Is it possible to perform a purely experimental analysis,
    \textit{i.e.}, excluding the theoretical points at negative $q^2$ and relying only on semileptonic and hadronic decays?
    \item The pseudo-observables obtained for the hadronic decays currently constrain only two relative phase between the three polarisations.
    What is the impact of a hypothetical theory determination of the absolute phase of the hadronic decays or an increased precision of the relative phases?
\end{itemize}

\begin{figure}[tbh!]
\includegraphics[width=.49\textwidth]{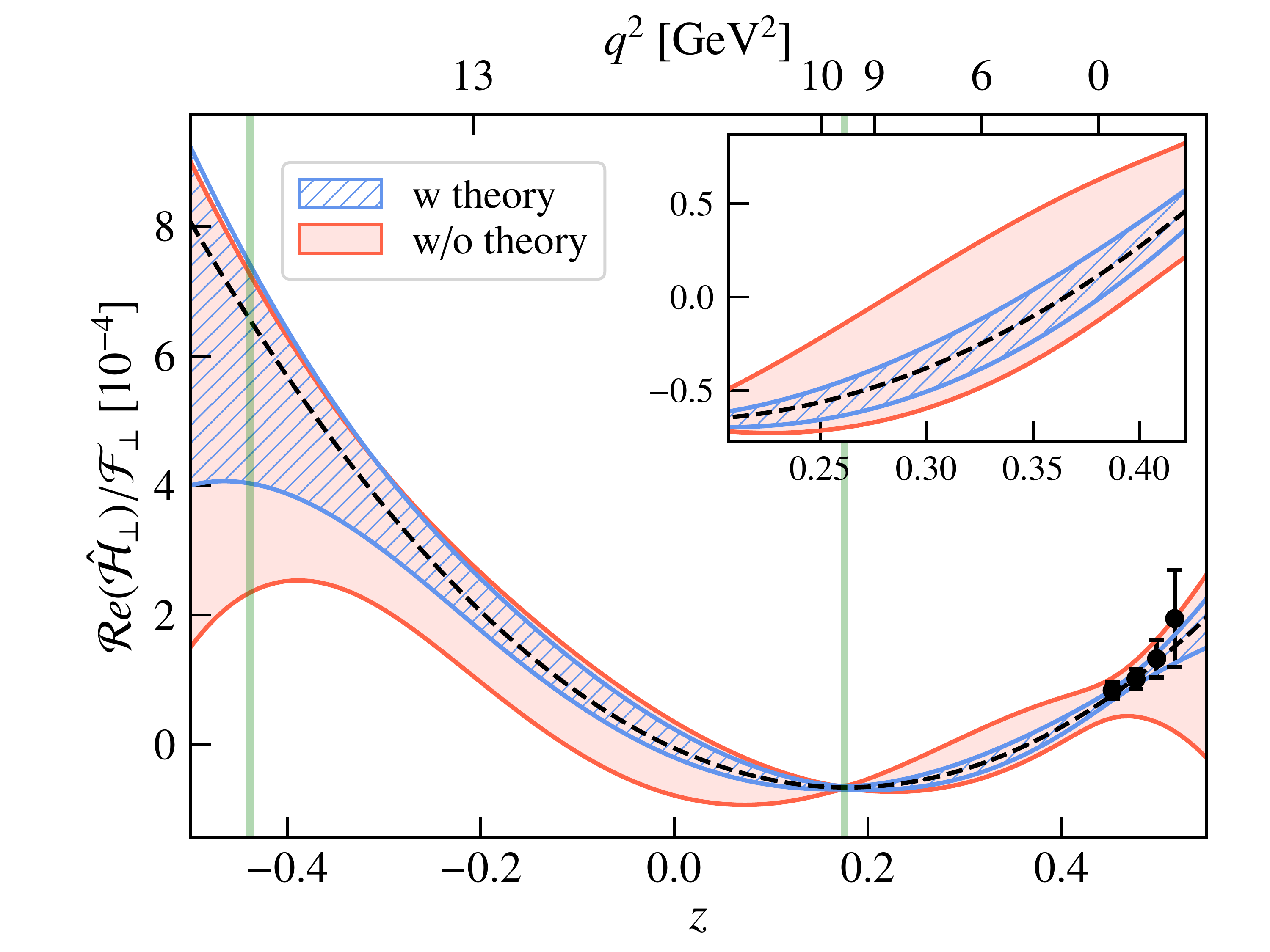} 
\includegraphics[width=.49\textwidth]{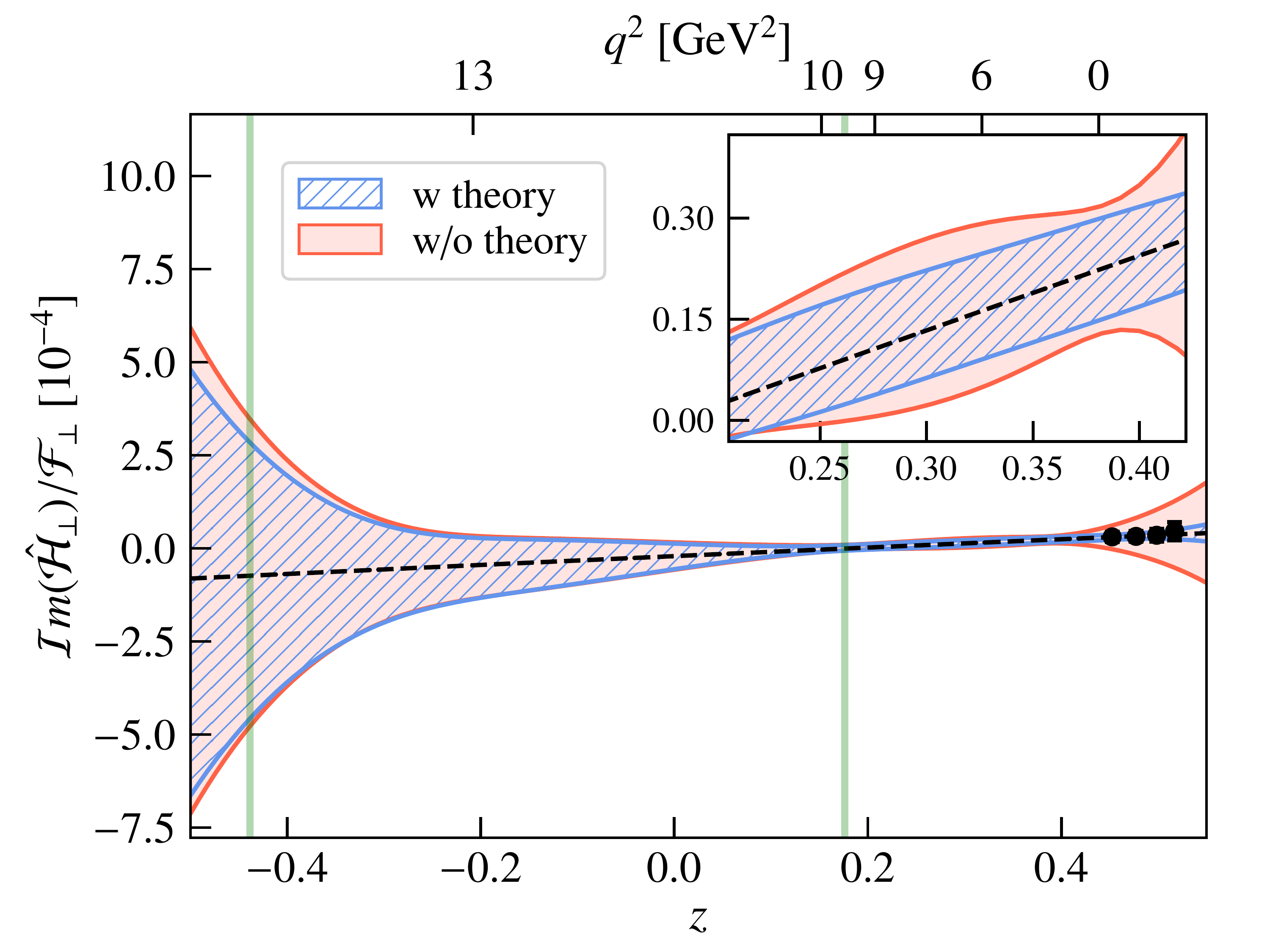} \\
\caption{%
    Results of the fits for the ratios $\hat{\mathcal{H}}_\perp(z)/\mathcal{F}_\perp(z)$ obtained with and without the
    theoretical points calculated at negative $q^2$. 
    Fits correspond to the BMP scenario, the expected statistics at LHCb Upgrade [50 fb$^{-1}$] and to $z^4$ truncation in the series expansion.
    The vertical bands correspond to the $J/\psi$ and $\psi(2S)$ regions; 
    the points to the theoretical inputs at negative $q^2$.
    The top right box of each plot zooms in the $q^2$ range between 1.1 and 9.0 GeV$^2$. 
    \label{fig:noTheo}
}
\end{figure}

To address the first point we repeat the analysis removing the constraints introduced by the theoretical calculation at negative $q^2$, and we examine the 
stability of the fit scanning different orders of the expansion.
We consider the BMP scenario corresponding to the expected statistics at LHCb Upgrade [50 fb$^{-1}$].
Fig.~\ref{fig:noTheo} shows the result of the fit assuming $z^4$ truncation of the expansion performed with and without the input from the theory points.
We find a strong model-bias, similarly to what is presented in section~\ref{sec:methodology-results:higher-orders}.
We conclude that a purely experimental analysis that combines information from the semileptonic decay and the hadronic $B\to K^*\lbrace J/\psi, \psi(2S)\rbrace$
decays is not currently possible.
The desired disentangling of the hadronic effect from possible NP contributions crucially relies on
the theory inputs from the points at negative $q^2$. \\

We investigate the benefits from hypothetical improvements to the constraints based on the hadronic decays.
First of all we note that, as shown in Fig.~\ref{fig:H}, the uncertainty on the hadronic correlator evaluated at the $J/\psi$
is extremely small. This is due to the fact that the region of the $J/\psi$ is highly constrained by the interference of theory
information at negative $q^2$ and the events of the semileptonic decay.
In fact we find that, already for datasets corresponding to the expected statistics at LHCb Run II, 
the impact of the pseudo-observables of the $J/\psi$ is negligible.
Furthermore, the fit is able to select the absolute phase of the $J/\psi$ with the same precision as the two relative phases.
As a consequence, in the following we focus on the impact of the $\psi(2S)$ pseudo-observables on the combined fit and 
we test whether it would be beneficial to have a measurement of the absolute phase of the $\psi(2S)$ and/or 
assuming future improvement in the measurement of the pseudo-observables of the $\psi(2S)$ 
(currently the two relative phases are weakly constrained~\cite{Bobeth:2017vxj}).
We proceed with a hypothetical constraint on the absolute phase of the $\psi(2S)$ and repeat the analysis 
with two configurations: first, we assume the relative uncertainty of the absolute phase of the $\psi(2S)$ to be similar to the relative phases' uncertainties. 
Second, we reduce the uncertainties of all phases of the $\psi(2S)$ to reflect the uncertainties of the $J/\psi$.
In both cases the central value of the absolute phase of the $\psi(2S)$ is set to the prediction obtained from the default set of 
parameters $\lbrace \alpha_\lambda^{(k)}\rbrace$ used for the production of the ensembles.

\begin{figure}[t!]
\includegraphics[width=.49\textwidth]{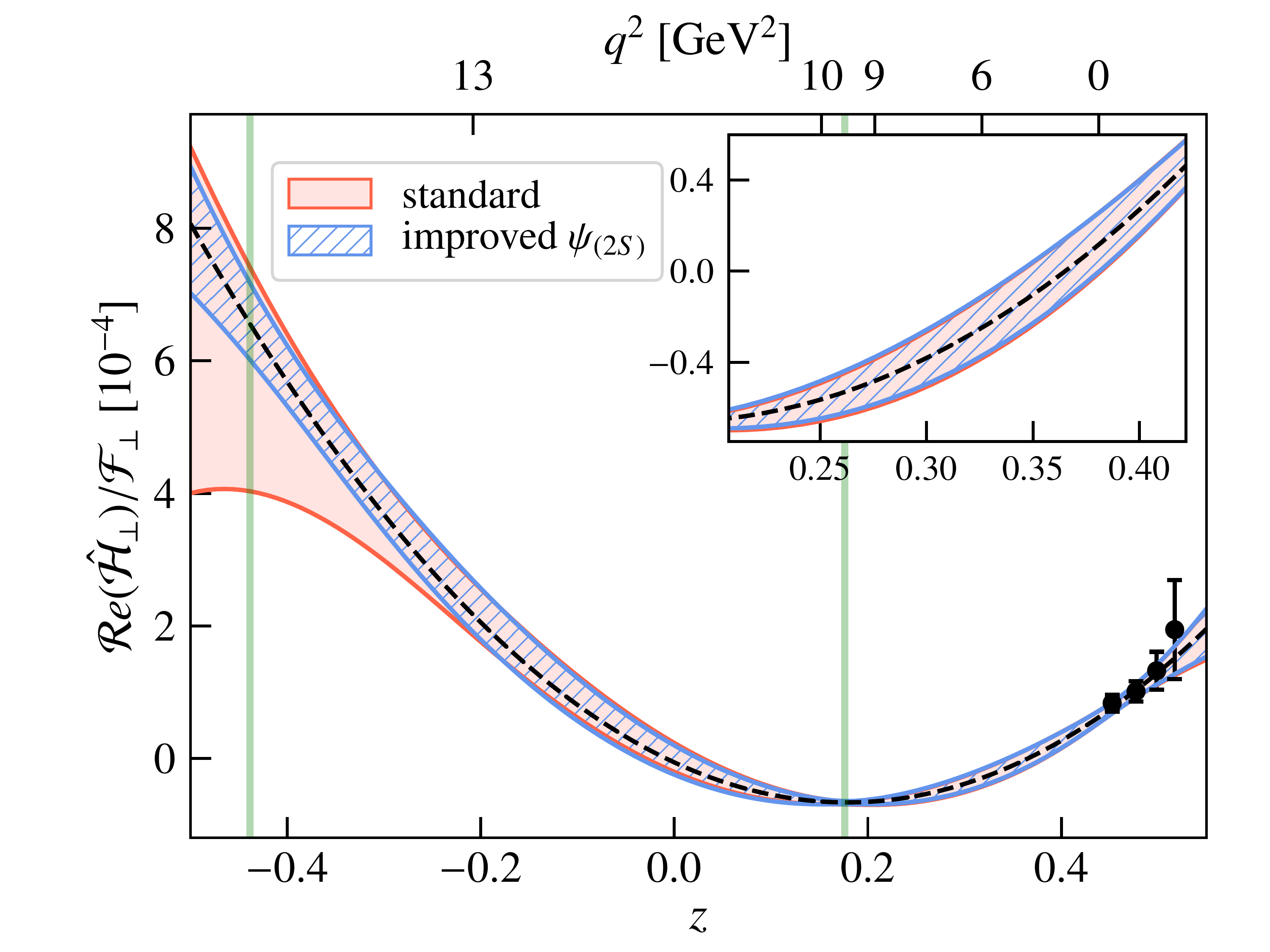} 
\includegraphics[width=.49\textwidth]{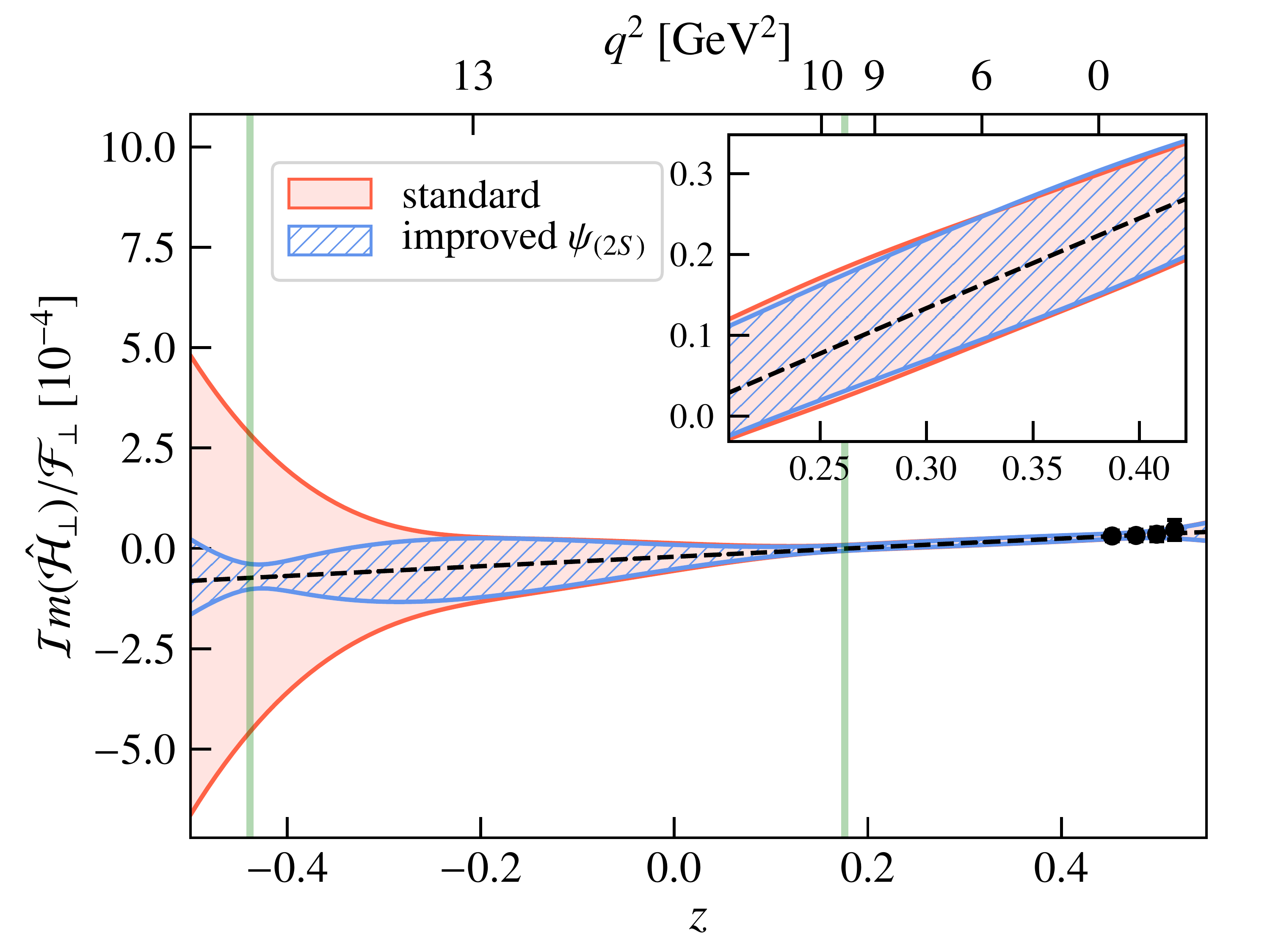} \\
\caption{%
    Results of the fits for the ratios $\hat{\mathcal{H}}_\perp(z)/\mathcal{F}_\perp(z)$ obtainded with the current status 
    of the theoretical and experimental knowledge and assuming future improvements on the $B\to K^*\psi(2S)$ measurements.
    Fits correspond to the BMP scenario, the expected statistics at LHCb Upgrade [50 fb$^{-1}$] and to $z^4$ truncation in the series expansion.
    The vertical bands correspond to the $J/\psi$ and $\psi(2S)$ regions; 
    the points to the theoretical inputs at negative $q^2$.
    The top right box of each plot zooms in the $q^2$ range between 1.1 and 9.0 GeV$^2$. 
    \label{fig:improvedPsi2S}
}
\end{figure}

We find that, even assuming the best case, the improvements on the determination of the WC $\mathcal{C}^{\textrm{NP}}_9$ are negligible.
Fig.~\ref{fig:improvedPsi2S} shows the comparison of the results obtained for the hadronic correlator for the analysis carried out 
with all the currently available information and the analysis that assumes the future improvements on the $\psi(2S)$ described above.
Both analyses are performed with datasets corresponding 
to the LHCb Upgrade [50 fb$^{-1}$] expected statistics and assume the $z^4$ truncation in the series expansion.
We find that the benefits produced by the assumed improvements on the $\psi(2S)$ pseudo-observables are limited 
to the region of the $\psi(2S)$.

\subsection{On the truncation of the series at $z^K$}

All the studies presented so far assumed a fixed set of initial values for the parameters $\lbrace \alpha_\lambda^{(k)}\rbrace$
in the production of the ensembles (obtained from~\cite{Bobeth:2017vxj}), in this section we investigate
the effect on the fit results of a different choice for the initial values of the hadronic parameters.
We investigate two options. First, we produce ensembles with non-zero coefficients for order 
of the expansion up to $z^3$ (\textit{i.e.} $\alpha^{(\lambda)}_{i\geq 4} = 0$).
Second, we produce ensembles with non-zero coefficients for order of the expansion up to 
$z^4$ (\textit{i.e.} $\alpha^{(\lambda)}_{i\geq 5} = 0$).

\begin{figure}[tbh!]
\includegraphics[width=.49\textwidth]{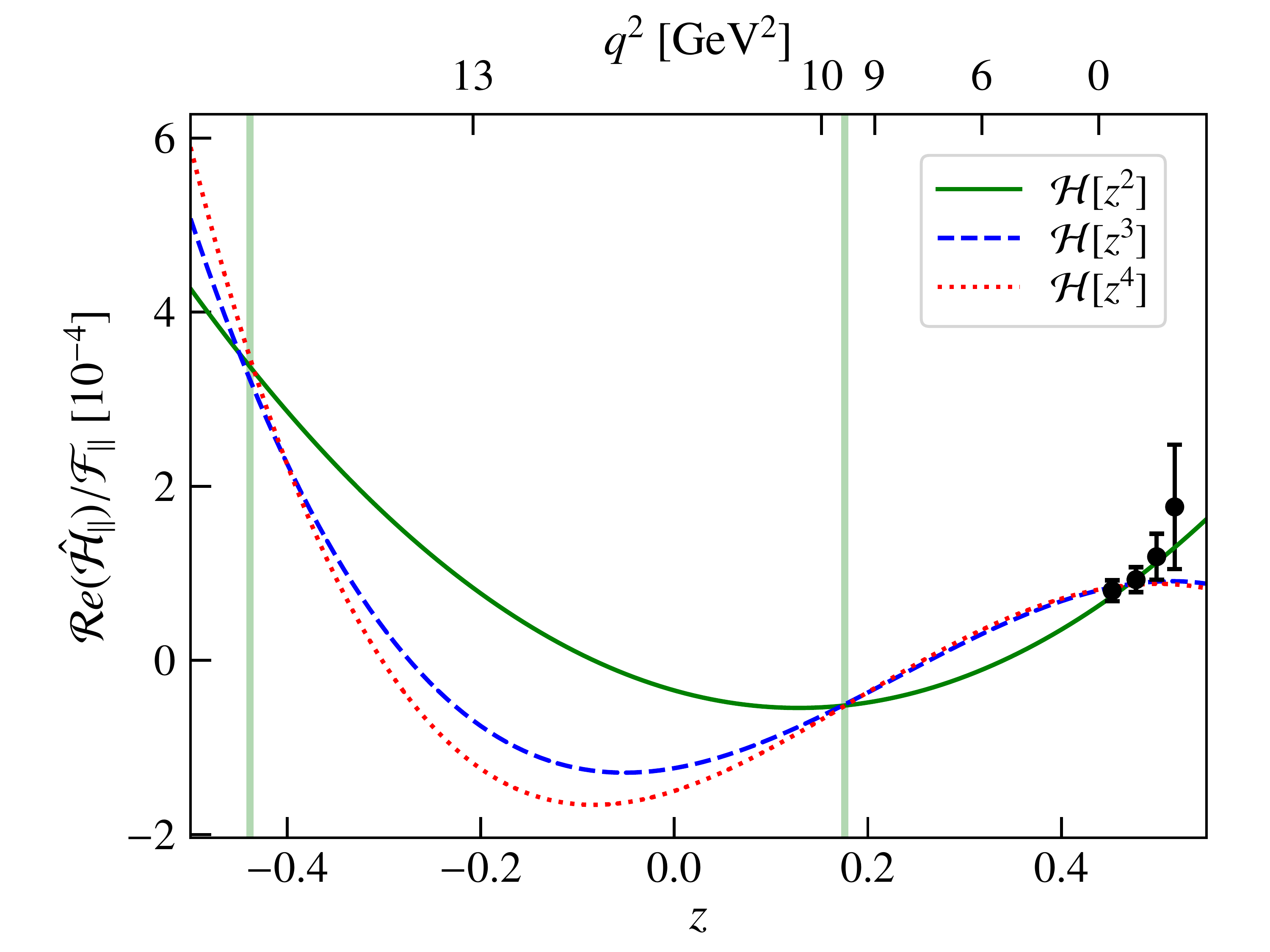} 
\includegraphics[width=.49\textwidth]{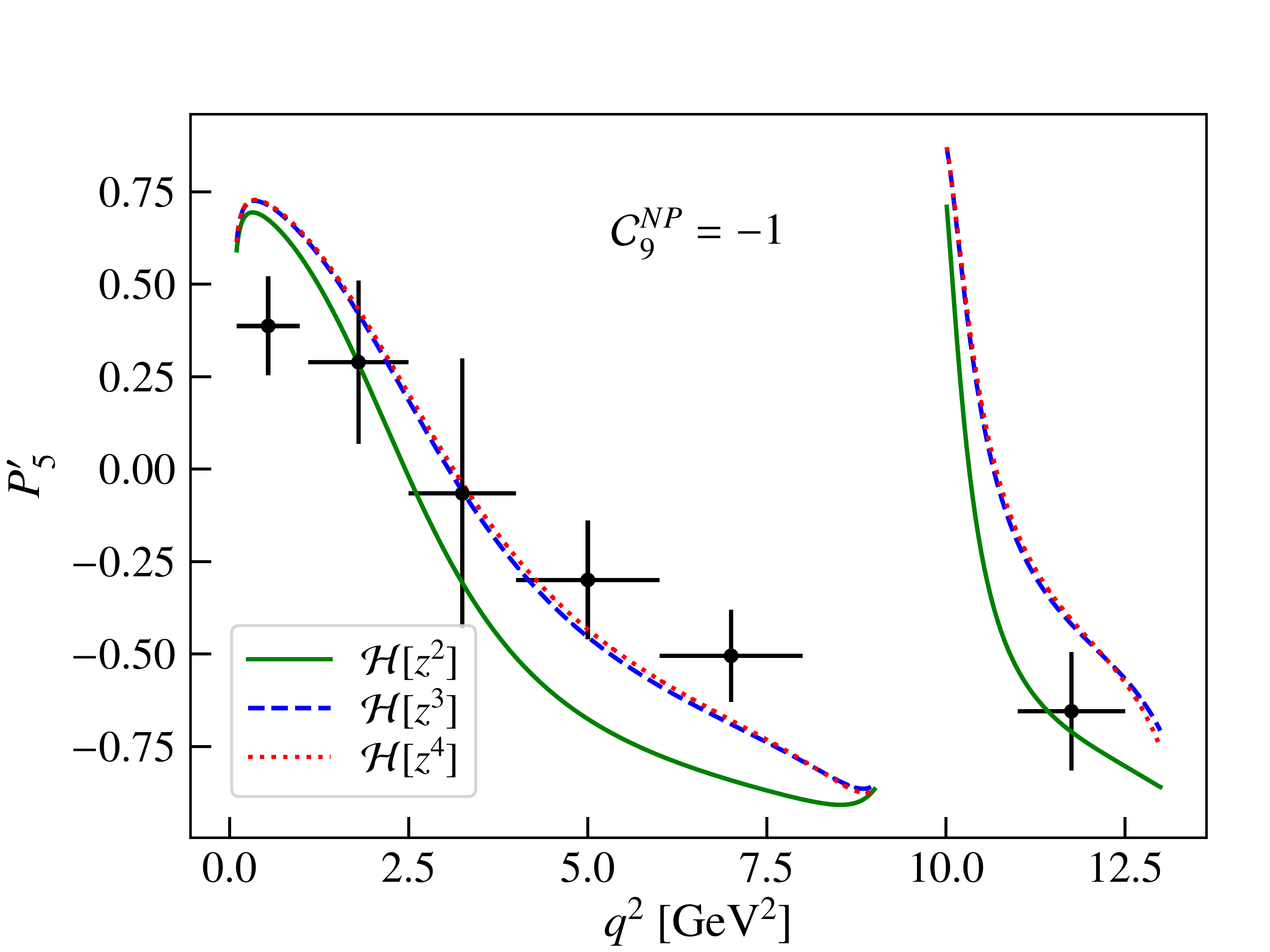} \\
\caption{%
    Left: Re$\hat{\mathcal{H}}_\parallel(z)/\mathcal{F}_\parallel(z)$ corresponding to the set of parameter 
    $\lbrace \alpha_\parallel^{(k)}\rbrace$ used for the production of the ensembles in the different hypotheses 
    as described in the text.
    Right: Projection of the different hypotheses in the $P'_5$ angular observable, all the three configurations assumes
    the BMP scenario. The result of the LHCb Run I analysis~\cite{Aaij:2013qta} is overlaid as reference.
    \label{fig:genZ3Z4}
}
\end{figure}

The choice of the above-mentioned non-zero coefficients is based on the following criteria:
they must be realistic (\textit{i.e.} compatible the theory predictions at negative $q^2$
and the pseudo-observables from the hadronic decays) and reduce the tension with the $P'_5$ anomaly
(\textit{i.e.} hadronic effects mimic the behaviour of NP).
The resulting set of parameters is shown in Fig.~\ref{fig:genZ3Z4} together with the value of the  $P'_5$
angular observable obtained in the different cases.

For each of the two generated configurations we perform the analysis as described in section~\ref{sec:CombinedAnalysis},
repeating the fit by varying the truncation of the expansion from $z^2$ to $z^5$.
Results are shown in Tables~\ref{tab:pullC9_genZ3} and~\ref{tab:pullC9_genZ4}, respectively.

Our conclusion can be summarised as follows:
\begin{itemize}
    \item fitting with the expansion truncated at $z^2$ (\textit{i.e.} lower order than what is used for the
    production of the ensembles)
    introduces a strong bias in the estimator for $\mathcal{C}^{\textrm{NP}}_9$;
    \item when the order of the truncation in the fitting procedure catches up the one used for the
    production of the ensembles
    the estimator for $\mathcal{C}^{\textrm{NP}}_9$ is unbiased;
    \item the uncertainty on $\mathcal{C}^{\textrm{NP}}_9$ varying the order of the fit follows the pattern observed in section~\ref{sec:CombinedAnalysis}.
	\item our lack of knowledge on the real description of the hadronic effects in nature can be investigated by scanning the order of the truncation of the series
	 until the central value of $\mathcal{C}^{\textrm{NP}}_9$ stabilises. Obviously, this procedure is bounded to the limit of the available statistics. 
\end{itemize}

\begin{table}
  \begin{center}
    \begin{tabular}{ c c c c c }
      \hline
      													  &      						 \multicolumn{2}{c}{LHCb Run2}									&  				   \multicolumn{2}{c}{LHCb Upgrade [50 fb$^{-1}$]}   				\\
      													  &      Re$\,\mathcal{C}_9^{\textnormal{NP}}$ mean      &      Re$\,\mathcal{C}_9^{\textnormal{NP}}$ sigma  &      Re$\,\mathcal{C}_9^{\textnormal{NP}}$ mean      &      Re$\,\mathcal{C}_9^{\textnormal{NP}}$ sigma    	 \\
      \hline
      $z^2$ fit    &     		-1.709 $\pm$ 0.007  		   &   		   0.138 $\pm$ 0.005    		&   		  -1.721 $\pm$ 0.003    		&    			0.060 $\pm$ 0.002      		\\ 
      $z^3$ fit    &    		 -1.004 $\pm$ 0.010  	  &      			 0.200 $\pm$ 0.007    	 &  		   -1.021 $\pm$ 0.005   		 &    			 0.106 $\pm$ 0.004     		 \\ 
      $z^4$ fit    &    		 -1.046 $\pm$ 0.011  	   &     			 0.214 $\pm$ 0.008     &    			-1.013 $\pm$ 0.006   	 	&     			 0.123 $\pm$ 0.004     	 	\\ 
      $z^5$ fit    &   		  -0.946 $\pm$ 0.013 	   &       		0.258 $\pm$ 0.009    	&    			 -0.986 $\pm$ 0.007    		 &     		0.144 $\pm$ 0.005     		 \\ 
      \hline
    \end{tabular}
    \caption{Expected central value and uncertainties for the $\mathcal{C}^{\textrm{NP}}_9$ observable obtained from $z^2$, $z^3$, $z^4$ and $z^5$ fits 
    for the BMP scenario when produced with non-zero $z^3$ coefficients as described in the text.}
    \label{tab:pullC9_genZ3}
  \end{center}
\end{table}

\begin{table}
  \begin{center}
    \begin{tabular}{ c c c c c }
      \hline
      													  &      						 \multicolumn{2}{c}{LHCb Run2}									&  				   \multicolumn{2}{c}{LHCb Upgrade [50 fb$^{-1}$]}   				\\
      													  &      Re$\,\mathcal{C}_9^{\textnormal{NP}}$ mean      &      Re$\,\mathcal{C}_9^{\textnormal{NP}}$ sigma  &      Re$\,\mathcal{C}_9^{\textnormal{NP}}$ mean      &      Re$\,\mathcal{C}_9^{\textnormal{NP}}$ sigma    	 \\
      \hline
      $z^2$ fit    &     		-1.813 $\pm$ 0.007  		   &   		   0.136 $\pm$ 0.005    		&   		  -1.824$\pm$ 0.003    		&    			0.063$\pm$ 0.002      		\\ 
      $z^3$ fit    &    		 -1.094 $\pm$ 0.010  	  &      			 0.196 $\pm$ 0.007    	 &  		   -1.188 $\pm$ 0.005   		 &    			 0.103 $\pm$ 0.004     		 \\ 
      $z^4$ fit    &    		 -1.049 $\pm$ 0.010  	   &     			 0.205 $\pm$ 0.007     &    			-1.018 $\pm$ 0.006   	 	&     			 0.119 $\pm$ 0.004     	 	\\ 
      $z^5$ fit    &   		  -0.938 $\pm$ 0.013 	   &       		0.257 $\pm$ 0.009    	&    			 -0.985 $\pm$ 0.007    		 &     		0.141 $\pm$ 0.005     		 \\ 
      \hline
    \end{tabular}
    \caption{Expected central value and uncertainties for the $\mathcal{C}^{\textrm{NP}}_9$ observable obtained from $z^2$, $z^3$, $z^4$ and $z^5$ fits 
    for the BMP scenario when produced with non-zero $z^4$ coefficients as described in the text.}
    \label{tab:pullC9_genZ4}
  \end{center}
\end{table}

\subsection{Simultaneous fit to $\mathcal{C}^{\textrm{NP}}_9$ and $\mathcal{C}^{\textrm{NP}}_{10}$}

\begin{figure}[tbh!]
\includegraphics[width=.44\textwidth]{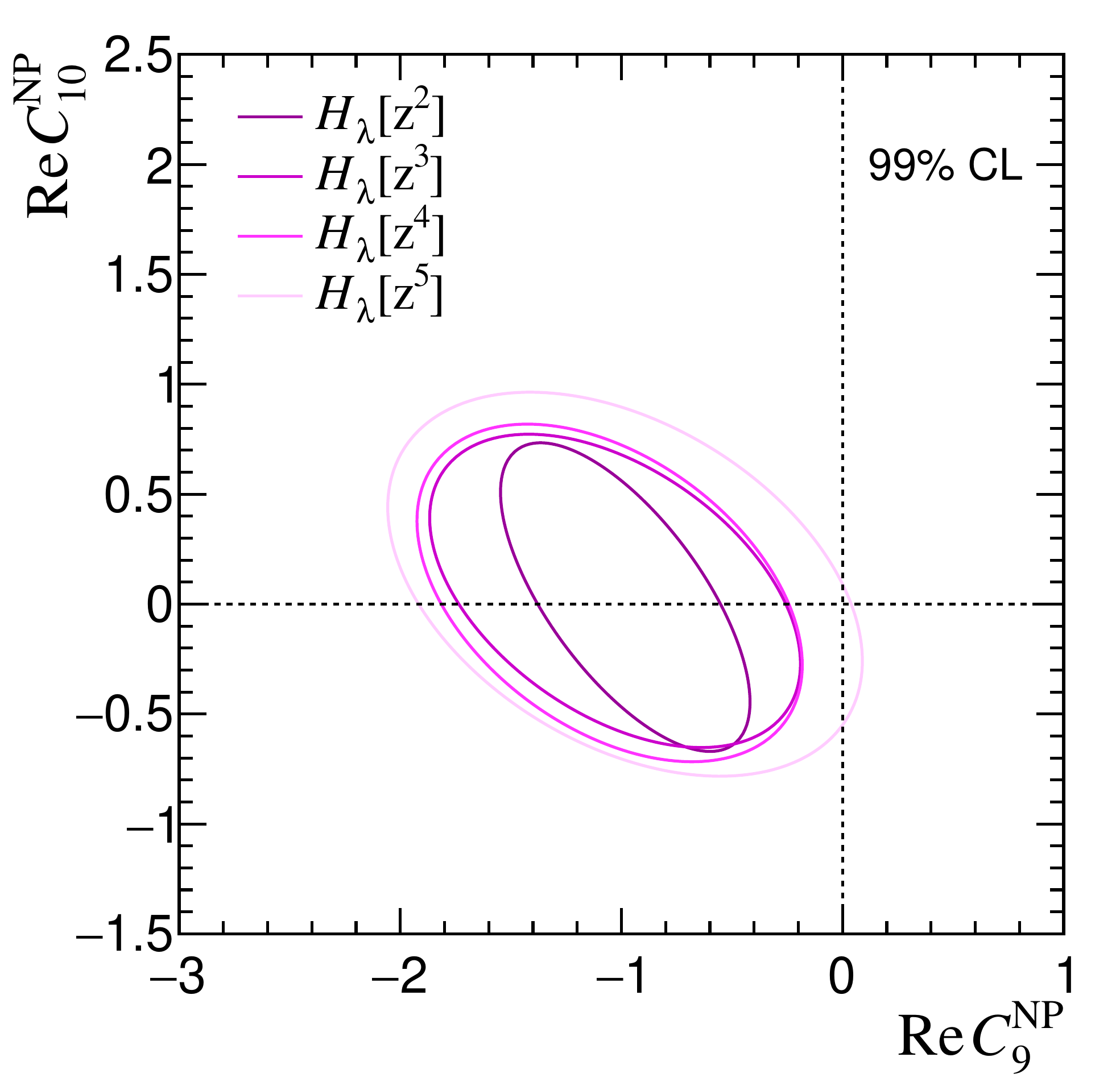} 
\includegraphics[width=.44\textwidth]{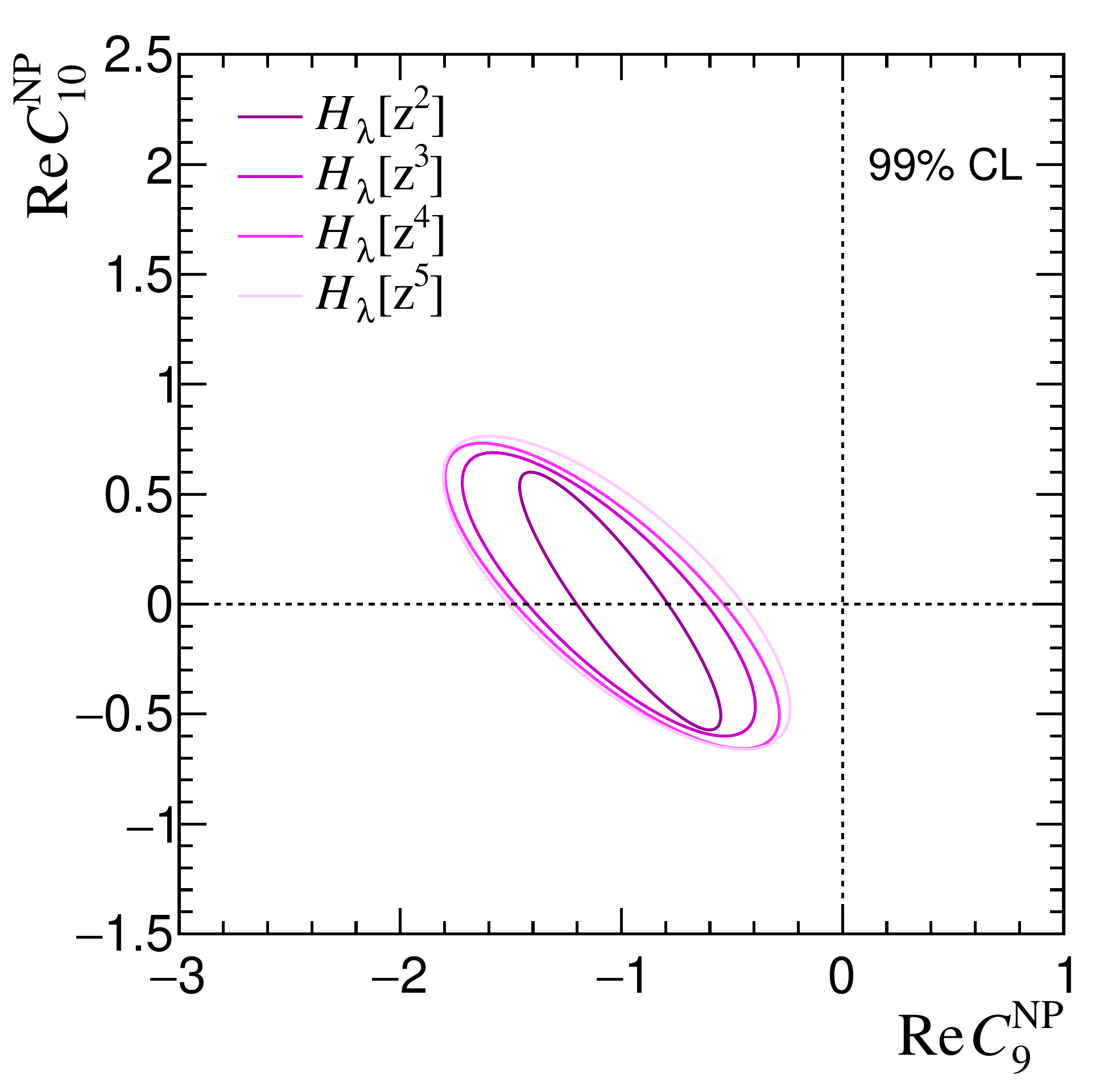} \\
\caption{%
    Two-dimensional sensitivity scans for the pair of Wilson coefficients $\mathcal{C}^{\textrm{NP}}_9$ and 
	$\mathcal{C}^{\textrm{NP}}_{10}$ for different non-local hadronic parametrisation models.
    The contours correspond to $3\,\sigma$ statistical-only uncertainty bands evaluated with
	the expected statistics after LHCb Run II (left) and LHCb Upgrade [50 fb$^{-1}$] (right).
    \label{fig:C9C10}
}
\end{figure}

\begin{figure}[tbh!]
\includegraphics[width=.44\textwidth]{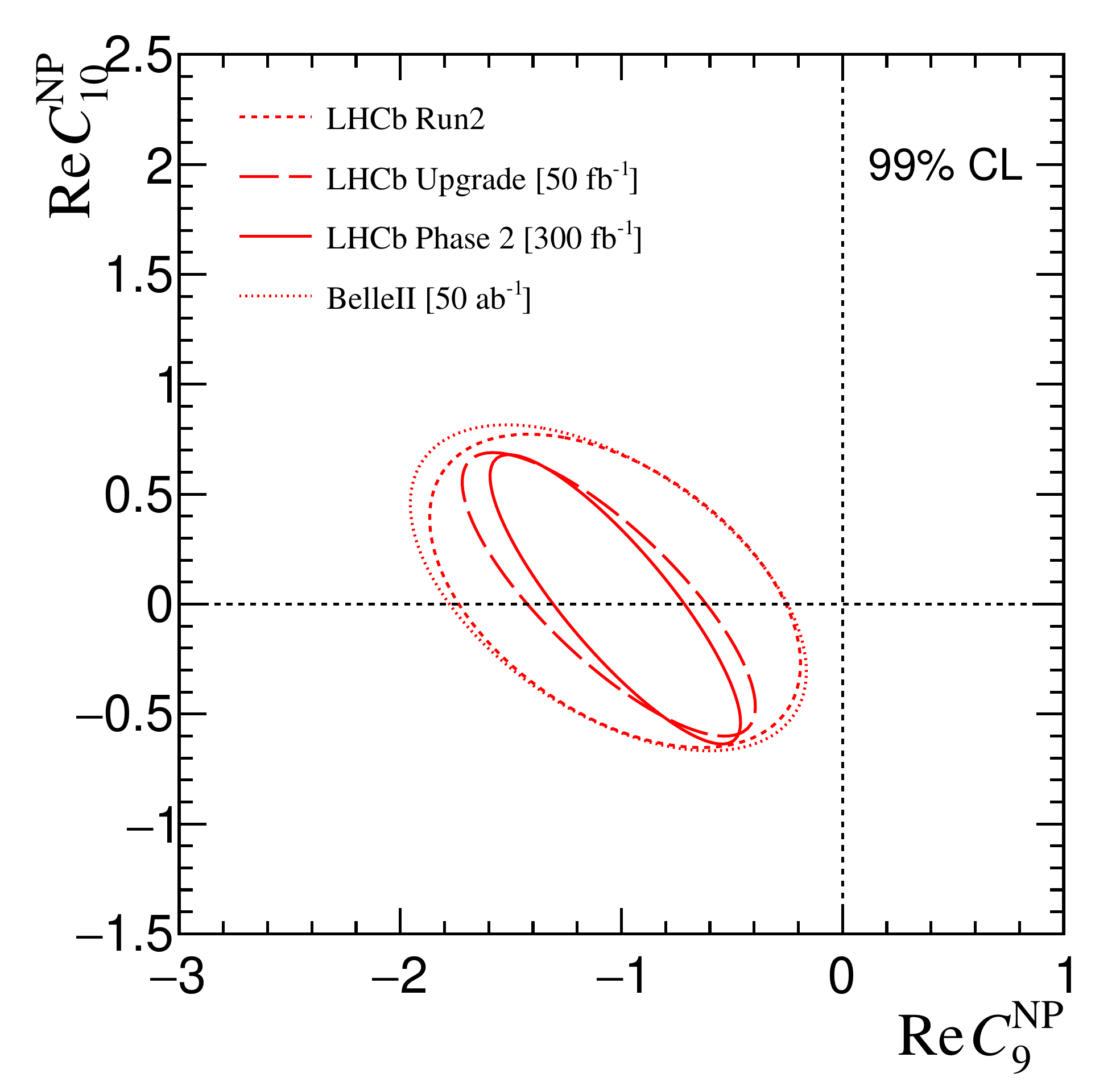} 
\caption{%
    Two-dimensional sensitivity scans for the pair of Wilson coefficients $\mathcal{C}^{\textrm{NP}}_9$ and 
	$\mathcal{C}^{\textrm{NP}}_{10}$ for the expected statistics corresponding to LHCb Run II (dotted), 
	LHCb Upgrade [50 fb$^{-1}$] (dashed), LHCb Upgrade [300 fb$^{-1}$] (solid) and Belle II [50 ab$^{-1}$] (long dashed).
    The contours correspond to $3\,\sigma$ statistical-only uncertainty bands obtained with $z^3$ fits.
    \label{fig:C9C10_Nev}
}
\end{figure}

As mentioned in section~\ref{sec:preliminaries}, $\mathcal{C}^{\textrm{NP}}_{10}$ does not suffer from pollution from hadronic non-local effects.
Nevertheless, it is interesting to extend the explored WCs parameter space and study the effect of floating $\mathcal{C}^{\textrm{NP}}_9$ and 
$\mathcal{C}^{\textrm{NP}}_{10}$ simultaneously in the fit.
We repeat the analysis as in section~\ref{sec:CombinedAnalysis}, producing 500 ensembles assuming the BMP scenario
(with NP inserted only in $\mathcal{C}^{\textrm{NP}}_9$, \textit{i.e.} $\mathcal{C}^{\textrm{NP}}_{10}=0$) with $z^2$ and performing the fit with 
truncation at the order $z^2$, $z^3$, $z^4$ and $z^5$.
The 2D pulls are shown in Fig.~\ref{fig:C9C10} while the single projections are reported in Tables~\ref{tab:pullC9C10_Run2} and 
\ref{tab:pullC9C10_Upgrade}.
Figure~\ref{fig:C9C10_Nev} shows the same result for different datasets corresponding to the expected statistics at LHCb Run II,
LHCb Upgrade [50 fb$^{-1}$ - 300 fb$^{-1}$] and Belle II [50 ab$^{-1}$]. For simplicity, only the results obtained from the $z^3$
analysis are shown.

We note that:
\begin{itemize}
    \item The uncertainty on $\mathcal{C}^{\textrm{NP}}_9$ varying the order of the fit follows the pattern observed in section~\ref{sec:CombinedAnalysis};
    \item due to the correlation between $\mathcal{C}^{\textrm{NP}}_9$ and $\mathcal{C}^{\textrm{NP}}_{10}$ the projection of the uncertainty on the single WC $\mathcal{C}^{\textrm{NP}}_9$ 
    is larger compared to the case of section~\ref{sec:CombinedAnalysis} when $\mathcal{C}^{\textrm{NP}}_{10}$ was fixed in the fit;
    \item besides the non-local hadronic effects, a precise determination of the WCs is limited by the uncertainties on the form factors,
    in particular, the precision on $\mathcal{C}^{\textrm{NP}}_{10}$ already saturates with the statistics expected to be collected at LHCb Run II 
    (see Fig.~\ref{fig:C9C10_Nev}). The precision on $\mathcal{C}^{\textrm{NP}}_{10}$ can be substantially improved by including constraints from 
    the decay $B_s \to \mu^+\mu^-$.
\end{itemize}

\begin{table}[t]
  \begin{center}
    \begin{tabular}{ c c c c c c }
      \hline
      \multicolumn{6}{c}{LHCb Run2}			\\
      													  &      Re$\,\mathcal{C}_9^{\textnormal{NP}}$ mean      &      Re$\,\mathcal{C}_9^{\textnormal{NP}}$ sigma  &      Re$\,\mathcal{C}_{10}^{\textnormal{NP}}$ mean      &      Re$\,\mathcal{C}_{10}^{\textnormal{NP}}$ sigma    	&    correlation   Re$\,\mathcal{C}_9^{\textnormal{NP}}$- Re$\,\mathcal{C}_{10}^{\textnormal{NP}}$\\
      \hline
      $z^2$ fit    &     	  -0.982 $\pm$ 0.008     		&       0.164 $\pm$ 0.006   		&      	 0.032 $\pm$ 0.010   		&       0.204 $\pm$ 0.007 	  &   -0.680 \\ 
      $z^3$ fit    &    	  -1.029 $\pm$ 0.012     		&       0.244 $\pm$ 0.009   		&      	 0.060 $\pm$ 0.010   		&       0.207 $\pm$ 0.007 	  &   -0.465 \\ 
      $z^4$ fit    &    	  -1.053 $\pm$ 0.013    		&       0.253 $\pm$ 0.009   		&      	 0.051 $\pm$ 0.011  			&       0.223 $\pm$ 0.008	  &   -0.427 \\ 
      $z^5$ fit    &   	 	 -0.983 $\pm$ 0.017     		&       0.312 $\pm$ 0.012  		&      	 0.091 $\pm$ 0.013	  		&       0.254 $\pm$ 0.009	  &   -0.400 \\ 
      \hline
    \end{tabular}
    \caption{Fit results obtained when floating $\mathcal{C}^{\textrm{NP}}_9$ and $\mathcal{C}^{\textrm{NP}}_{10}$ for $z^2$, $z^3$, $z^4$ and $z^5$ fits. 
    Ensembles are produced for the BMP scenario with the corresponding statistics expected at the LHCb Run II.}
    \label{tab:pullC9C10_Run2}
  \end{center}
\end{table}

\begin{table}[t]
  \begin{center}
    \begin{tabular}{ c c c c c c }
      \hline
      \multicolumn{6}{c}{LHCb Upgrade [50 fb$^{-1}$]}			\\
      													  &      Re$\,\mathcal{C}_9^{\textnormal{NP}}$ mean      &      Re$\,\mathcal{C}_9^{\textnormal{NP}}$ sigma  &      Re$\,\mathcal{C}_{10}^{\textnormal{NP}}$ mean      &      Re$\,\mathcal{C}_{10}^{\textnormal{NP}}$ sigma    	&    correlation   Re$\,\mathcal{C}_9^{\textnormal{NP}}$- Re$\,\mathcal{C}_{10}^{\textnormal{NP}}$\\
      \hline
      $z^2$ fit    &     	  -1.005 $\pm$ 0.007    		&      0.132 $\pm$ 0.005  		&      	0.014 $\pm$ 0.008  		&      0.171 $\pm$ 0.006		  &      -0.891 \\ 
      $z^3$ fit    &    	  -1.057 $\pm$ 0.010    		&      0.193 $\pm$ 0.007  		&      	0.044 $\pm$ 0.009  		&      0.188 $\pm$ 	0.007	 	 &       -0.791 \\ 
      $z^4$ fit    &    	  -1.041 $\pm$ 0.011    		&      0.220 $\pm$ 0.008  		&      	0.037 $\pm$ 0.010  		&      0.202 $\pm$ 0.007		  &      -0.781 \\ 
      $z^5$ fit    &   		  -1.021 $\pm$ 0.011    		&      0.228 $\pm$ 0.008  		&      	 0.051 $\pm$ 0.010  		&      0.207 $\pm$ 0.007		  &       -0.734 \\ 
      \hline
    \end{tabular}
    \caption{Fit results obtained when floating $\mathcal{C}^{\textrm{NP}}_9$ and $\mathcal{C}^{\textrm{NP}}_{10}$ for $z^2$, $z^3$, $z^4$ and $z^5$ fits. 
    Ensembles are produced for the BMP scenario with the corresponding statistics expected at the LHCb Upgrade [50 fb$^{-1}$].}
    \label{tab:pullC9C10_Upgrade}
  \end{center}
\end{table}

\subsection{Unbinned determination of angular observables}

\begin{figure}[tbh!]
\includegraphics[width=.49\textwidth]{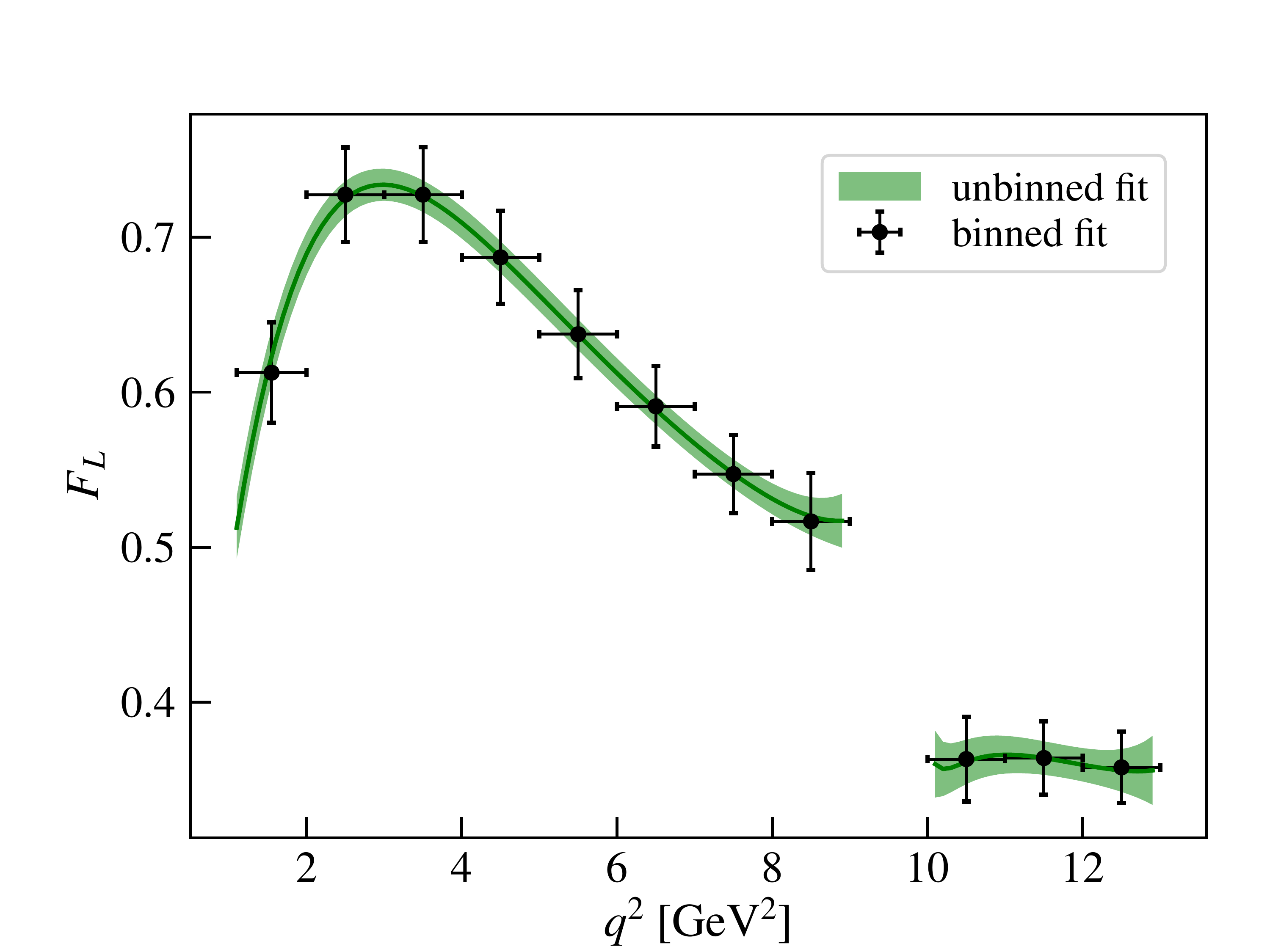} 
\includegraphics[width=.49\textwidth]{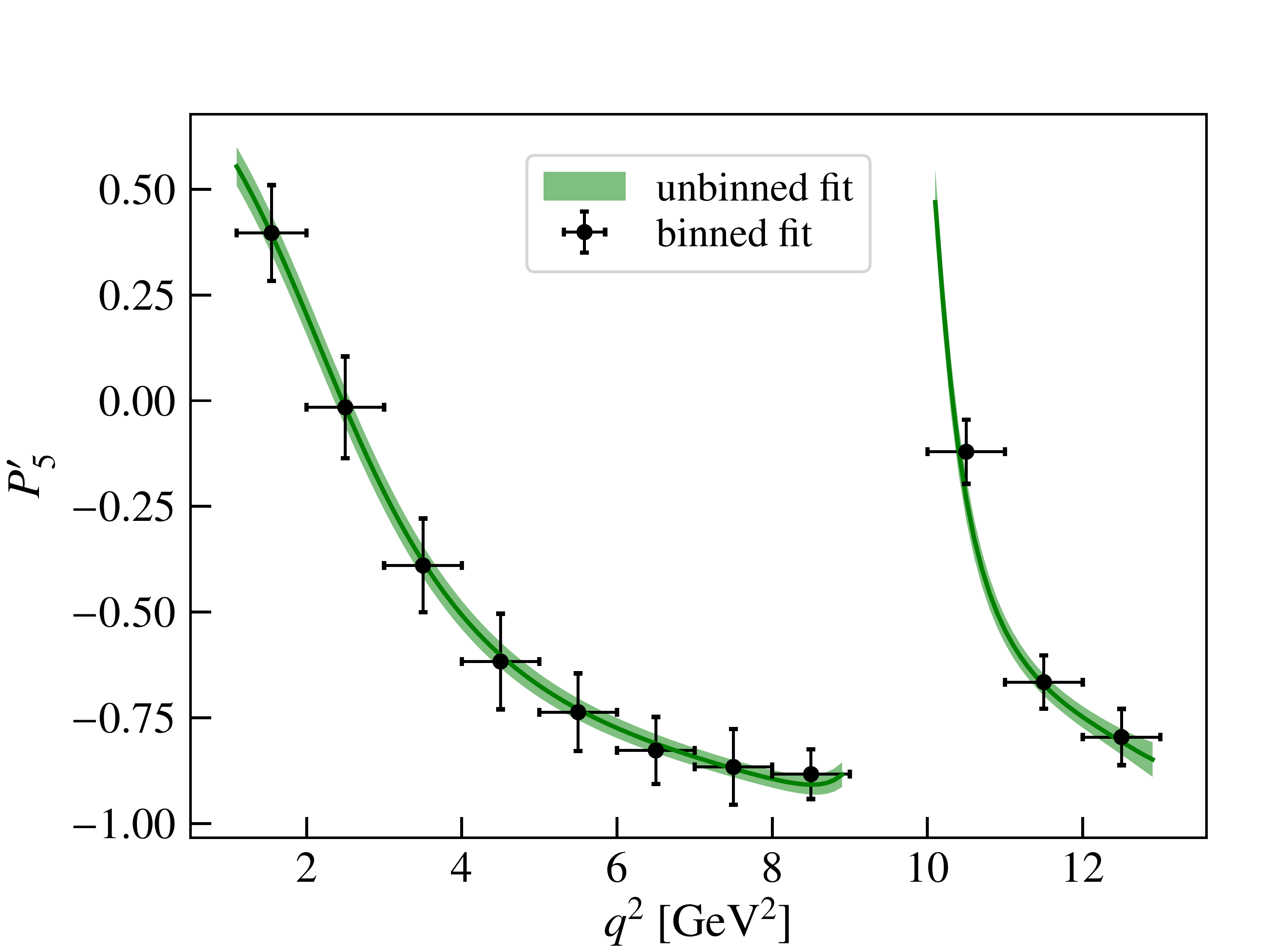} \\
\caption{%
    Angular observables $F_L$ and $P'^5$ obtained a-posteriori from the unbinned amplitude fit results
    compared with the binned angular analysis. Both approaches analyse the same set of ensembles generated 
    with the BMP scenario and the expected statistics at LHCb Run II.
    \label{fig:FLandP5p}
}
\end{figure}

One of the benefits of our proposed approach is that it takes advantage of the full unbinned description of the decay
and, additionally, the amplitude fit allows to reproduce confidence intervals for the commonly used angular observables.
In the following we investigate the statistical uncertainty expected for the obtained angular observables.
We perform the analysis on 500 ensembles generated with the expected statistics at LHCb Run II
and we repeat the fit with different truncations at the $z^2$, $z^3$, $z^4$ and $z^5$ order.
We find that the uncertainty on all the angular observables is independent on the assumption on the
truncation of the series expansion, leading to clean results free from systematic uncertainties on the hadronic parametrisation.
It is interesting to compare the statistical uncertainty on the angular observables obtained by the unbinned 
amplitude fit with respect to the binned approach.
We perform a binned fit on the same ensembles generated above, splitting the datasets in 1 GeV$^2$ $q^2$ bins,
and we fit the obtained angular distributions with the signal PDF $\dd^3\Gamma / \dd^3 \Omega$ as described in~\cite{Aaij:2015oid}.
Figure~\ref{fig:FLandP5p} shows the improvement on the statistical uncertainty on the angular observables
obtained by the unbinned amplitude fit compared to a $q^2$ binned angular analysis.

\section{Conclusion}

Measurements of angular observables in the decay $B\to K^*\mu^+\mu^-$
have shown discrepancies with respect to Standard Model predictions,
mainly in the angular observable known as $P_5^{\prime}$. This
anomaly has been widely discussed in the literature, in particular
since non-local charm contributions are challenging to be predicted
from a theory point of view. 
Here we carried out a sensitivity study of the decay $B\to K^*\mu^+\mu^-$, 
taking care to account for hadronic matrix
elements of both local and non-local operators in a model-independent fashion.
This is done by performing an extended unbinned maximum likelihood fit, 
which allows to use the full information of the data and extract simultaneously the
hadronic parameters and the Wilson Coefficients. 
Other studies have proposed methods to disentangle NP from non-local hadronic 
effects in (un)binned likelihood fits.
Following the parametrisation of Ref.~\cite{Bobeth:2017vxj} we studied the properties of the fit for 
a large number of different  scenarios, by using simulated pseudoexperiments. 
Fitting with the order of the $z$-expansion used in the production of the ensembles 
leads to an unbiased value of $\mathcal{C}_9^{\textnormal{NP}}$. 
Increasing the order of the $z$-expansion, the central value of $\mathcal{C}_9^{\textnormal{NP}}$
stays unbiased while the uncertainty increases. It is observed that theory constraints strongly 
mitigate this problem. 
The increase in the uncertainty on $\mathcal{C}_9^{\textnormal{NP}}$ is roughly 
one order of magnitude smaller than the statistical uncertainty obtained adding one 
order to the $z$-expansion in the fit.
The fact that the uncertainty on $\mathcal{C}_9^{\textnormal{NP}}$ steadily 
increases with the order of the polynomial of the $z$-expansion does not allow us 
to rigorously assign a systematic uncertainty due to the truncation of the $z$-expansion 
a-priori. 
We also found that the unbinned fit allows to extract additional information on 
the non-local matrix elements from semi-muonic decay events alone. 
Our study goes beyond previous works, and assesses in a quantitative way the 
model-dependency due to our ansatz for the non-local hadronic contributions. 
Our approach allows systematic  improvements (through increasing the truncation 
order in $z$) and estimation of systematic uncertainties (through varying 
the truncation order even when the data is described well).
In addition, the unbinned fit can be used for the determination of the 
usual angular observables with precision beyond what can be expected with the 
standard binned approach. We find that the angular observables obtained with this
method do not exhibit any sizeable model bias due to the truncation of the $z$-expansion.
It should be emphasised that the gain in sensitivity cannot be directly read 
from~Fig.~\ref{fig:FLandP5p} since the points of the binned likelihood fit are all uncorrelated. 
To fully access the comparison of the two approaches, a binned fit to the angular observables 
with the same model should be performed.

While not strictly necessary, improving our knowledge of the $B\to K^*\psi(2S)$ amplitudes can give us greater confidence
of the obtained fit results. We therefore encourage revisiting their analysis at the present $B$-physics experiments.
The application of the unbinned fit for the decays $B \to K \mu^+\mu^-$ and $\Lambda_b \to \Lambda \mu^+ \mu^-$
has not been studied here. 
It is unclear how the lack of phase information on the $J/\psi$ and $\psi(2S)$ poles will affect the efficiency
of the unbinned fit.
We therefore encourage dedicated sensitivity studies for these decays as a natural extension of our present work.

\acknowledgments

We are grateful to Christoph Bobeth, S\'ebastien Descotes-Genon, Patrick Owen,
and Javier Virto for useful discussions. We also thank Christoph Bobeth and Javier Virto
for valuable comments on the manuscript.
D.v.D.~gratefully acknowledges partial support by the Swiss National Science
Foundation (SNF) under contract 200021-159720, and by the Emmy Noether
programme of the Deutsche Forschungsgemeinschaft (DFG) under grant DY 130/1-1.
A.M., N.S.~and R.S.C.~also acknowledge the support by SNF under contracts 173104 and 174182.

\bibliography{references}

\end{document}